\documentclass{aa}  
\usepackage{graphicx}
\usepackage{txfonts}
\begin{document}
   \title{The low-mass YSO CB230-A: investigating the protostar and its jet 
   with NIR spectroscopy and Spitzer observations.}

   \author{F. Massi\inst{1} \and C. Codella\inst{2} \and J. Brand\inst{3}
          \and L. Di Fabrizio\inst{4} \and J.G.A. Wouterloot\inst{5}}

   \offprints{F. Massi (fmassi@arcetri.astro.it)}

\institute{INAF-Osservatorio Astrofisico di
Arcetri, Largo E. Fermi 5, 50125 Firenze, Italy \and
INAF, Istituto di Radioastronomia, Sezione di
Firenze, Largo E. Fermi 5, 50125 Firenze, Italy \and
INAF, Istituto di Radioastronomia, Via P. Gobetti 101, 40129 
Bologna, Italy \and
INAF, Telescopio Nazionale Galileo, 38700 Santa Cruz de la Palma, Spain \and
Joint Astronomy Centre, 660 N. A'ohoku Place, University Park,
Hilo, 96720 HI, USA}

   \date{Received date; accepted date}


\titlerunning{NIR spectroscopy of the low-mass CB230-A YSO and its jet}
\authorrunning{Massi et al.}

  \abstract
{
To investigate the 
earliest phases of star formation and study how newly-born stars 
interact with the surrounding medium, we performed
a line and continuum survey at NIR and mm-wavelengths of a 
sample of relatively isolated Bok globules. 
}
{We present a follow-up observational program of a star-forming site
in the globule CB230. From narrow-band continuum observations of this site, we 
had discovered a bright [FeII] jet, which originates in the low-mass YSO 
CB230-A. We aim to investigate the physical properties of the region 
from where the jet is launched.
}
{Our analysis was carried out
using low-resolution NIR spectra acquired with the
camera NICS at the TNG telescope, with $JH$ and $HK$ grisms and a 1 
arcsec-wide slit.
These observational data were complemented with infrared photometric data from the 
Spitzer space telescope archive.
}
{
The relevant physical properties of CB230-A were constrained by SED 
fitting of fluxes from the NIR to the mm. 
The YSO spectrum exhibits a significant number of atomic and
molecular emission lines and absorption features.
The characteristics of this
spectrum suggest that we are observing a region in the close vicinity of
CB230-A, i. e.  its photosphere and/or an active accretion disk.
The spectra of the knots in the jet contain a large number of
emission lines, including a rich set of [FeII] lines. 
Emission due to H$_2$ and [FeII] 
are not spatially correlated, confirming that [FeII] and H$_2$ are excited by 
different mechanisms, in agreement with the models where [FeII] 
traces dissociative J-shocks and molecular hydrogen traces 
slower C-shocks. By using 
intensity ratios involving density-sensitive [FeII] lines, 
we estimated the electron densities along the jet to be 
$6 \times 10^3$--$1 \times 10^4$ cm$^{-3}$. This indicates 
either high density post-shock regions of ionised gas or regions with a high 
degree of ionisation. 
}
{By combining the present data with previously obtained maps at NIR- and 
mm-wavelengths, the emerging scenario is that CB230-A is a Class 0/I YSO 
driving an atomic jet that  is observed to be almost monopolar probably due 
to its 
inclination to the plane of the sky and the resulting higher extinction of its 
red side. This primary jet appears to be sufficiently energetic to open the cavity 
visible in the NIR images and drive the large-scale molecular outflow 
observed at mm-wavelengths. 
{{\bf
CB230-A was revealed to be a good location to 
test the innermost structure of accreting low-mass protostars.
}}
}


\maketitle
%

\section{Introduction}
\label{intro}
High-velocity jets driven by collimated winds generated by Young Stellar
Objects (YSOs) interact strongly with the surrounding medium, cleaning up the 
high-density material hosting the star-forming process. Jets are also thought 
to remove angular momentum from the accreting matter allowing lower angular 
momentum gas to carry on the build-up process of a new star.
In addition, when they travel into the ambient cloud, jets create shocks that 
heat (up to thousands of K) and compress the gas, and drive bipolar molecular 
outflows on larger spatial scales containing colder (10-100 K) swept-up
material. As a consequence, the study of the acceleration and
collimation of jets is fundamental to understanding star formation.

\begin{figure}
\centering
\includegraphics[angle=0,width=8.5cm]{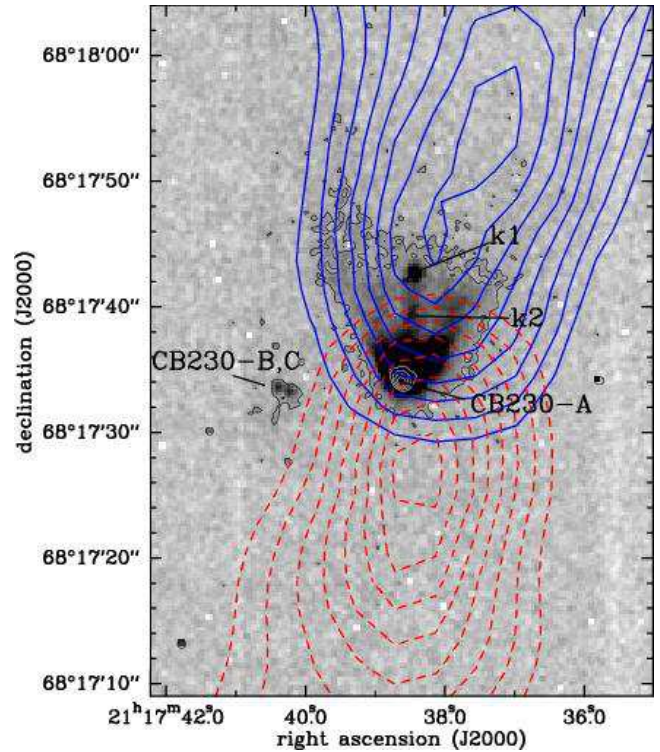}
\caption{Continuum-subtracted [FeII] map (grey scale; Massi et al.\ 2004) of 
CB230: the white contours mark the brightest emission indicating the 
YSO CB230-A. The label CB230-B,C mark two NIR objects, that 
could make a triple system with CB230-A. The labels k1 and k2 indicate the
two [FeII] knots defining the jet. Also visible is an extended diffuse 
component showing 
the cavity opened by the YSO mass ejection. In contours, the
CO(3--2) integrated emission of the blue (continuous) and red (dashed) lobes 
of the molecular outflow driven by CB230-A (Brand et al.\ 2008; JCMT-data). The 
integration intervals are from $-6$ to $+1.37$ km s$^{-1}$ (blue lobe)
and from $+4.68$ to $+14$ km s$^{-1}$ (red lobe). Contours range from 3 to 20
K km s$^{-1}$ in steps of 1 K km s$^{-1}$. The blue lobe extends for 
$\sim 120\arcsec$, the red one is roughly half as long.}
\label{Fig:Out}
\end{figure}

Some of the most accessible places to survey jets are represented by Bok globules
(Bok \& Reilly \cite{bok}), which are relatively isolated molecular clouds
mainly associated with low-mass star formation (e.g. Huard et al.
\cite{huard99} and references therein). The globule CB230 (Clemens \& 
Barvainis \cite{clemens})
is a good example, given the 
simultaneous presence of almost all the ingredients expected in a typical star 
formation scenario: YSOs, a hot atomic jet, and a colder molecular outflow. 
One of the YSOs in the cloud (hereafter called CB230-A) was 
detected in the Near-Infrared (NIR) as well as at sub-millimetre wavelengths
(Yun \& Clemens \cite{yuncle92, yuncle94a}; Launhardt \& Henning
\cite{launhardt97}; Huard et al. \cite{huard99}; Young et al. \cite{young}). 
It is also associated with a bipolar outflow, observed in CO (Yun \& Clemens 
\cite{yuncle94b}). 
Using N$_{2}$H$^{+}$ observations, Chen et al.\ (\cite{chen}) found a 
clear velocity gradient across the molecular core of 
$\sim 8.8$~km\,s$^{-1}$\, pc$^{-1}$, in the same direction 
as the line 
connecting CB230-A with a source $\sim 10 \arcsec$ east. 
This latter source close to CB230-A was 
detected at 7~${\mu}$m with ISOCAM and coincides with a possibly double
NIR source (CB230-B,C; see Fig.~\ref{Fig:Out}).  

The distance to CB230 was discussed by Launhardt \& Henning 
(\cite{launhardt97}), who placed it at 450~pc.
This was derived by
associating CB230 with a larger molecular cloud complex (Cepheus Flare),
based on their respective radial velocities. The authors estimated that the
given distance is accurate to within 30\%.
Kun (\cite{kun}) discussed the distance to the Cepheus Flare
molecular cloud complex and for L1177 (the dark cloud
hosting CB230) quotes a distance in the range 300--560~pc, where the
lower value is the kinematic distance and the higher one is
the distance to the nearby (but probably unrelated)
reflection nebula vdB141. Although kinematic distances lower than 
1 kpc are of doubtful reliability, the lowest value agrees with the
distance ($300 \pm 30$~pc) derived for the dark clouds in the
area by studying the cumulative distribution of stellar
distance moduli (Wolf diagram). Young et al. (\cite{yy06}) assumed
$d=288$~pc taking this value from Straizys et al.\ (\cite{straizys}), 
who derived the distance to a group of dark clouds (close
to L1177, but not including it) using photometry of 79
stars. In particular, L1167/L1174 were found to lie at
$d = 288 \pm 25$~pc, in agreement with the lower value
indicated by Kun (\cite{kun}) for L1177.
In summary, it seems that $d = 450$~pc ($\pm$ 30\%) is a robust determination
and 288~pc may be assumed as a safe lower limit.

CB230-A is classified as a Class~0/Class~I source (e.\ g., Froebrich
\cite{froeb}). Estimates of the bolometric luminosity, based on the IRAS
fluxes, range from 7.7~L$_{\sun}$ (Froebrich \cite{froeb}) to
12~L$_{\sun}$ (Launhardt \& Henning \cite{launhardt97}), both for a 
distance of 450~pc, i. e.  the distance we adopt.
From our sub-mm observations with SCUBA (Brand et al.\ \cite{brand}),
we derive a gas mass of 3.4~M$_{\sun}$, which agrees within a factor of 2 with
the mass estimated by Launhardt (\cite{launhardt01}) from observations at
1.3~mm with the IRAM 30-m. Launhardt (\cite{launhardt01}) also found a disk of
mass $\sim 0.1$~M$_{\sun}$ around the protostar, based on observations
at 1mm continuum with the Owens Valley Radio Interferometer.
By using three different
evolutionary models, Froebrich (\cite{froeb}) derived an age in the range
$1.7 \times 10^{4}$--$2.2 \times 10^{5}$~yrs and a final stellar mass in the 
range 0.3--0.9~M$_{\sun}$.

In the framework of our ongoing multi-frequency study of 
star formation in Bok globules (Massi et al.\ \cite{ma2004}; Codella et al.\ 
\cite{codella}), we carried out narrow-band NIR observations centred on the
1.644~$\mu$m [FeII] and 2.122~$\mu$m H$_{\rm 2}$ lines.
As a result, we detected a bright [FeII] jet in CB230, which (i)
originates in CB230-A and extends for $\sim$ 0.02 pc, (ii) is directed 
along the same direction as the molecular outflows axis (N-S), and (iii) is 
defined by two knots (called k1 and k2) superimposed on fainter elongated 
emission observed also in H$_{\rm 2}$ (Massi et al.\ \cite{ma2004}).
Figure~\ref{Fig:Out} illustrates this by displaying an overlay of 
the [FeII] image and the CO(3--2) bipolar outflow as mapped by
Brand et al. (\cite{brand}).
These [FeII] and H$_{\rm 2}$ NIR lines are particularly useful, since [FeII]
emission traces fast and dissociative J-shocks, which should outline the inner
jet-channel, whereas H$_{\rm 2}$ emission is expected to trace slower C-shocks 
that probably originate in bow-shocks, probing the region of interaction between the 
jet and the ambient gas.
The CB230 NIR maps therefore support a complex scenario, which, to be confirmed,
requires further [FeII] 
and H$_{\rm 2}$ spectroscopic NIR observations; 
these data were obtained and enabled us
to constrain the physical
properties of the gas (such as extinction and excitation temperature) in the 
region where the jet interacts with the surrounding dense gas close to its origin, 
i. e. close to the central system, which consists of a protostar and, presumably, an
accretion disk.


\section{Observations and data reduction}
The spectroscopic observations were carried out on the night of August 3, 2004,
with the NIR camera NICS at the Telescopio Nazionale Galileo (TNG) at La 
Palma. We obtained spectra through the
grisms $JH$ ($1.15-1.75$~$\mu$m) and $HK$ ($1.40-2.50$~$\mu$m), using a 
$1\arcsec$--wide slit. This provided a spectral resolution of R$\sim$500. The slit 
was aligned along the [FeII] jet defined by the k1 and k2 knots, 
hence 
encompassing also the driving source (the YSO). 
Six spectra of the target were
obtained through the grism $JH$ in an ABBAAB set of cycles (i.\ e.
integrating when the source is in a position A on the detector, then nodding 
the telescope along the slit projection on the sky to move the source to a 
different position B on the detector, integrating again and reversing the 
cycle, etc.). The integration time for each spectrum was 900~s. Two spectra 
were taken 
through the grism $HK$ in a single AB cycle, with integration times of 600~s 
each. The star HIP97033 (spectral type GO~V) was observed in between the 
$JH$- and the $HK$-cycles on the source, by acquiring four spectra in an ABBA cycle with 2~s of 
integration each, through both the $JH$- and the $HK$-grism.
Other spectra of this star and of HIP108772 (spectral type O9~V)
were taken at the beginning of the night, but they could not be used to correct 
the target spectra for telluric absorption because they were observed
at least 5 hours before the target. However, they were used to obtain
the intrinsic spectrum of HIP97033 and perform other tests on the telluric 
correction. The weather conditions did not remain stable and observations 
of the target could be carried out only in the late night due to clouds.

Data reduction 
was done
using IRAF\footnote{IRAF is distributed by the National Optical Astronomy Observatories,
    which are operated by the Association of Universities for Research
    in Astronomy, Inc., under cooperative agreement with the National
    Science Foundation.}. 
All frames were flat-field corrected
by using spectra of a halogen lamp, following the prescriptions given on the 
TNG web page, i. e. the halogen frames were normalised row-by-row by dividing by a 
low-degree polynomial fit of the background. All frames were 
wavelength-calibrated and rectified by using reference spectra of an Argon 
lamp. Then, each pair of subsequent frames (i. e. the A and B exposures in 
AB- or BA-cycles) were subtracted to remove background emission and other 
biases. Since a single frame (i. e. A or B in AB- or BA-cycles) is an image 
where the source spectrum appears as a row of bright pixels, after subtraction
a frame is obtained that contains two spectra of the source, one positive and one 
negative with respect to the average background level.
By careful examination of the [FeII] images in Massi et al.\
(\cite{ma2004}), we estimated the extent of the two emission knots and the YSO
in the spatial direction, and extracted their spectra from each subtracted 
frame (two per frame). Two spectra of HIP97033 per subtracted frame were also 
extracted.

Using the solar spectrum as an input, a synthetic spectrum was constructed
(following Maiolino et al.~\cite{maio}) that represented an arbitrarily-scaled 
version of the intrinsic spectrum of HIP97033 (assuming a radial velocity of 
$-21.6$ km s$^{-1}$ and a rotational velocity of 10 km s$^{-1}$, as listed in 
the SIMBAD database).
%
%
For all extracted spectra of HIP97033, the wavelength calibration 
was then further 
refined by aligning the intrinsic absorption lines in each stellar 
spectrum
with the corresponding ones in the synthetic (reference) spectrum. This is 
required because moving the telescope produces small shifts in the wavelength
scale of each frame (i. e. there exists a shift between the wavelength scale
in the Ar-lamp frames and that in each of the source frames). 
Finally, all other spectra were further wavelength-corrected
(to minimize this instrumental effect) by shifting their wavelength 
scale to align their atmospheric absorption lines with those in the final 
stellar spectra.
This correction translated into shifts of $\sim 10^{-3}$ $\mu$m
(i. e. $\sim 10$ $\AA$, independent of wavelength) 
both for the telluric standard and the targets.
All spectra of a given source were then coadded to obtain the
highest signal-to-noise ratio. The atmospheric transmission was corrected
for as explained in Maiolino et al.\ (\cite{maio}), i. e., by dividing the 
spectra of the 
knots and the YSO by that of the telluric standard and then 
multiplying them
by the synthetic intrinsic spectrum of the star obtained from the solar 
spectrum.

\subsection{Flux calibration}

We first dereddened the spectra over the entire wavelength range
by adopting the reddening law of Rieke \& Lebofsky (\cite{r&l}).
We tentatively derived $A_{V}$ from [FeII] line ratios (see 
Sect.~\ref{emi:lin}; the uncertainties affecting the obtained values
are further discussed in Sect.~\ref{red:scatt}). Once $A_{V}$ was known, the 
reddening correction was obtained by dividing each spectrum by a curve 
proportional to the extinction as a function of the wavelength. Then, the 
dereddened spectra were flux-calibrated according to the following 
prescriptions. For the knots
(i. e. Figs.~\ref{Fig:Speck1} and \ref{Fig:Speck2}),
we required that the total flux of the 
[FeII] line at 1.645~$\mu$m equalled the narrow-band photometric 
determinations of Massi et al.\ (\cite{ma2004}). 

For CB230-A
(i. e. Fig.~\ref{Fig:Specyso}), 
each of the $J, H$, and $K$ segments of the dereddened YSO 
spectrum was first multiplied by the total transmission of the 2MASS system
(atmosphere excluded); the derived flux densities were then integrated over each 
of the $J, H$, and $K$ passbands and divided by the integral 
of the total 2MASS transmission over the corresponding passband.
The resulting mean flux densities were eventually set to be equal to the flux 
densities derived from the 2MASS magnitudes of the YSO 
(2MASS21173862+6817340, $J=14.085 \pm 0.062$, $H=11.712$, $K=10.487$) after 
being dereddened using the same $A_{V}$ used to deredden the entire
spectrum. We then obtained two conversion factors (counts to flux density) 
from $JH$ (one from $J$ and one from $H$) and two conversion factors from 
$HK$ (one from $H$ and one from $K$).

Obviously, the two conversion factors for each of the two YSO spectra (i. e.
$JH$ and $HK$) should coincide. For the $HK$ spectrum, the conversion 
factors 
obtained from $H$ and $K$ 
agree to within a few percent, whereas 
for the $JH$ spectrum, those obtained from $J$ and $H$ differ 
by $\sim 30$\%. 
The two different calibration factors within $JH$ probably reflect 
errors in the 
photometry due to the presence of diffuse emission towards the YSO and possible 
time-variability slightly affecting the SED (spectra and 2MASS data 
were obviously not taken simultaneously). 

However, the calibrated YSO spectra have different 
continuum emission slopes within the 
overlapping $H$ interval, with differences in flux of up to $\sim 20$ \%.
The same problem is found when calibrating with the integrated flux
in the [FeII] line at 1.645~$\mu$m, which produces differences in the slope of the
underlying continuum towards the knots of up to $\sim 30$ \%.
Inaccuracies in the telluric correction are discussed in Appendix~\ref{ap:a}
and do not appear to affect the spectra significantly.
The most likely cause is the intense, extended emission around the source,
which could have prevented an accurate placement of the slit on the central source.  

\subsection{Spitzer observations}

Spitzer observations of CB230 were taken from the
Spitzer public archive. They consisted of IRAC images at
the four wavelengths (3.6, 4.5, 5.8, and 8~$\mu$m) and MIPS
images at 24 and 70~$\mu$m. 
The observations were part of the GTO program 124, ``IRAC and MIPS Imaging 
and IRS Spectroscopy of Pre and Post Main Sequence Stellar Systems'' 
(Principal Investigator G. Gehrz). Overviews of IRAC and MIPS were given by 
Fazio et al.\ (\cite{fazio1}) and Rieke et al.\ (\cite{rieke04}), respectively.

We retrieved all post-BCD (Basic Calibrated Data) images, which were produced by 
the pipeline version S14.0.0 (IRAC), S16.1.0 (MIPS at 24~$\mu$m), 
and S16.0.1 (MIPS at 70~$\mu$m). All MIPS data were obtained in the 
photometric
small field modes (fine scale at 70~$\mu$m). After checking all images for 
possible problems as described in the instrument Data Handbooks
\footnote{http://ssc.spitzer.caltech.edu/}, we decided not to
repeat the reduction steps starting from the BCD images, and 
photometry was carried out 
on the post-BCD images.

CB230-A and CB230-B(,C) are resolved in the IRAC and MIPS 24~$\mu$m images, 
but blended in the MIPS 70~$\mu$m images. CB230-B and CB230-C are 
unresolved even in
the IRAC images, so they were considered to be a single source
(hereafter, CB230-B+C). 
Aperture photometry of both CB230-A and CB230-B+C 
was performed using DAOPHOT in IRAF. We selected aperture radii of 2 pixels
($\sim 2 \arcsec$ [IRAC], $3 \arcsec$ [MIPS 24~$\mu$m], and $35 \arcsec$ 
[MIPS 70~$\mu$m]). The background contribution was measured within annuli
with inner radius of 2 pixels and width of 4 pixels (IRAC),
inner radius of $6\arcsec$ and width of $7 \arcsec$ (MIPS at 24~$\mu$m),
and inner radius of $39 \arcsec$ and width of $26 \arcsec$ (MIPS at 24~$\mu$m).
We adopted aperture corrections given on the
Spitzer web page\footnote{http://ssc.spitzer.caltech.edu/}, for the used apertures 
and sky annuli. In the case of the MIPS
24~$\mu$m image, the aperture photometry for each of the two sources was performed 
after
subtraction of the other source by PSF fitting. The photometric uncertainties
were computed as described by Reach 
et al.\ (\cite{reach}) and Gordon et al.\ (\cite{gordon}).
We note that due 
to flux density nonlinearities, aperture photometry is expected 
to underestimate the MIPS 70~$\mu$m flux, probably by $\sim 20$ \% (Gordon et 
al.\ \cite{gordon}). 
The fluxes were converted into magnitudes in the IRAC (Reach et 
al.\ \cite{reach}) and MIPS (see the MIPS handbook) systems and are listed in 
Table~\ref{spitz:tab}.

\begin{table}
\caption[]{Spitzer magnitudes of CB230-A and CB230-B+C.
\label{spitz:tab}}
\begin{tabular}{c c c}
\hline
Band & CB230-A & CB230-B+C \\
($\mu$m) & (mag) & (mag) \\
\hline
$3.6$ & $9.101 \pm 0.004$ & $12.62 \pm 0.04$ \\
$4.5$ & $8.537 \pm 0.006$ & $12.06 \pm 0.04$ \\
$5.8$ & $8.033 \pm 0.006$ & $11.49 \pm 0.06$ \\ 
$8$ & $7.88 \pm 0.01$ & $10.21 \pm 0.01$ \\
$24$ & $2.847 \pm 0.004$ & $4.48 \pm 0.02$ \\
$70^{a}$ & \multicolumn{2}{c}{$-3.75 \pm 0.07$} \\ 
\hline
\end{tabular}

$^a$ CB230-A and CB230-B+C are spatially
unresolved in this band and appear as a single source.
\end{table}

\section{Results} 

\subsection{Emission lines}
\label{emi:lin}

\begin{table*}
\caption[]{Emission lines observed towards the knots k1 and k2 and
the YSO CB230-A: detections and tentative detections. The values between
brackets are the measured r.m.s. \label{emlines}}
\begin{tabular}{ccccc}
\hline
   &      & k1$^a$ & k2$^a$ & CB230-A$^a$ \\
Transition & $\lambda$$^b$ ($\mu$m)  & \multicolumn{3}{c}{$F$ (10$^{-15}$ erg cm$^{-2}$ s$^{-1}$)} \\
\hline
[FeII] a$^4$ D$_{\rm 7/2}$ -- a$^6$ D$_{\rm 9/2}$ & 1.2567 & 78.6(2.0) & 125.8(4.8) & 38.7(2.3) \\
$[$FeII] a$^4$ D$_{\rm 1/2}$ -- a$^6$ D$_{\rm 1/2}$ & 1.2704$^c$ & 9.5(2.1)$^c$ & -- & 7.9(1.6)$^c$ \\
$[$FeII] a$^4$ D$_{\rm 3/2}$ -- a$^6$ D$_{\rm 3/2}$ & 1.2788 & 10.8(2.0) & -- & 14.2(1.6) \\
$[$FeII] a$^4$ D$_{\rm 5/2}$ -- a$^6$ D$_{\rm 5/2}$ & 1.2943 & 11.1(2.9) & -- & 7.7(2.0) \\
$[$FeII] a$^4$ D$_{\rm 7/2}$ -- a$^6$ D$_{\rm 7/2}$ & 1.3206 & 22.7(2.4) & 21.6(5.1) & 16.0(1.6)  \\
$[$FeII] a$^4$ D$_{\rm 5/2}$ -- a$^6$ D$_{\rm 3/2}$ & 1.3278 & 7.6(2.3) & -- & -- \\
$[$FeII] a$^2$ I$_{\rm 11/2}$ -- b$^2$ H$_{\rm 9/2}$ & 1.5246 & 5.0(1.3) & 10.7(5.2)$^d$ & -- \\
$[$FeII] a$^4$ D$_{\rm 5/2}$ -- a$^4$ F$_{\rm 9/2}$ & 1.5335 & 12.3(2.0) & 11.2(5.2)$^d$ & 21.5(5.9) \\
$[$FeII] a$^4$ D$_{\rm 3/2}$ -- a$^4$ F$_{\rm 7/2}$ & 1.5995 & 5.7(1.7) & -- & -- \\
$[$FeII] a$^4$ D$_{\rm 7/2}$ -- a$^4$ F$_{\rm 9/2}$ & 1.6436 & 58.4(1.1) & 60.8(2.8) & 46.5(6.1) \\
$[$FeII] a$^4$ D$_{\rm 1/2}$ -- a$^4$ F$_{\rm 5/2}$ & 1.6638 & 5.2(1.7) & -- & -- \\
$[$FeII] a$^4$ D$_{\rm 5/2}$ -- a$^4$ F$_{\rm 7/2}$ & 1.6769$^e$ & 4.7(1.1) & -- & 10.9(6.3)$^e$ \\
HI 11-4                                             & 1.6811$^e$ &  --      & -- & 15.7(6.3)$^e$ \\
H$_2$ 1--0 S(1)                                     & 2.122  & 11.7(3.1) & 15.4(2.5) & 38.3(2.9) \\
HI 7-4                                             & 2.1661 &  --      & -- & 10.8(2.9) \\
\hline
\end{tabular}

$^a$ The $JHK$ spectra have been dereddened using the visual extinctions derived
from the [FeII] 1.645/1.321 line ratio (see Tab.~\ref{ext:lr}).
$^b$ The wavelengths of the [FeII] lines are taken from the NIST atomic spectra database.
$^c$ Blended with a telluric line.
$^d$ S/N $\simeq$ 2: tentative detection.
$^e$ The HI and [FeII] lines at 1.68~$\mu$m are blended.
\end{table*}

\begin{figure*}
\centering
\includegraphics[width=16cm]{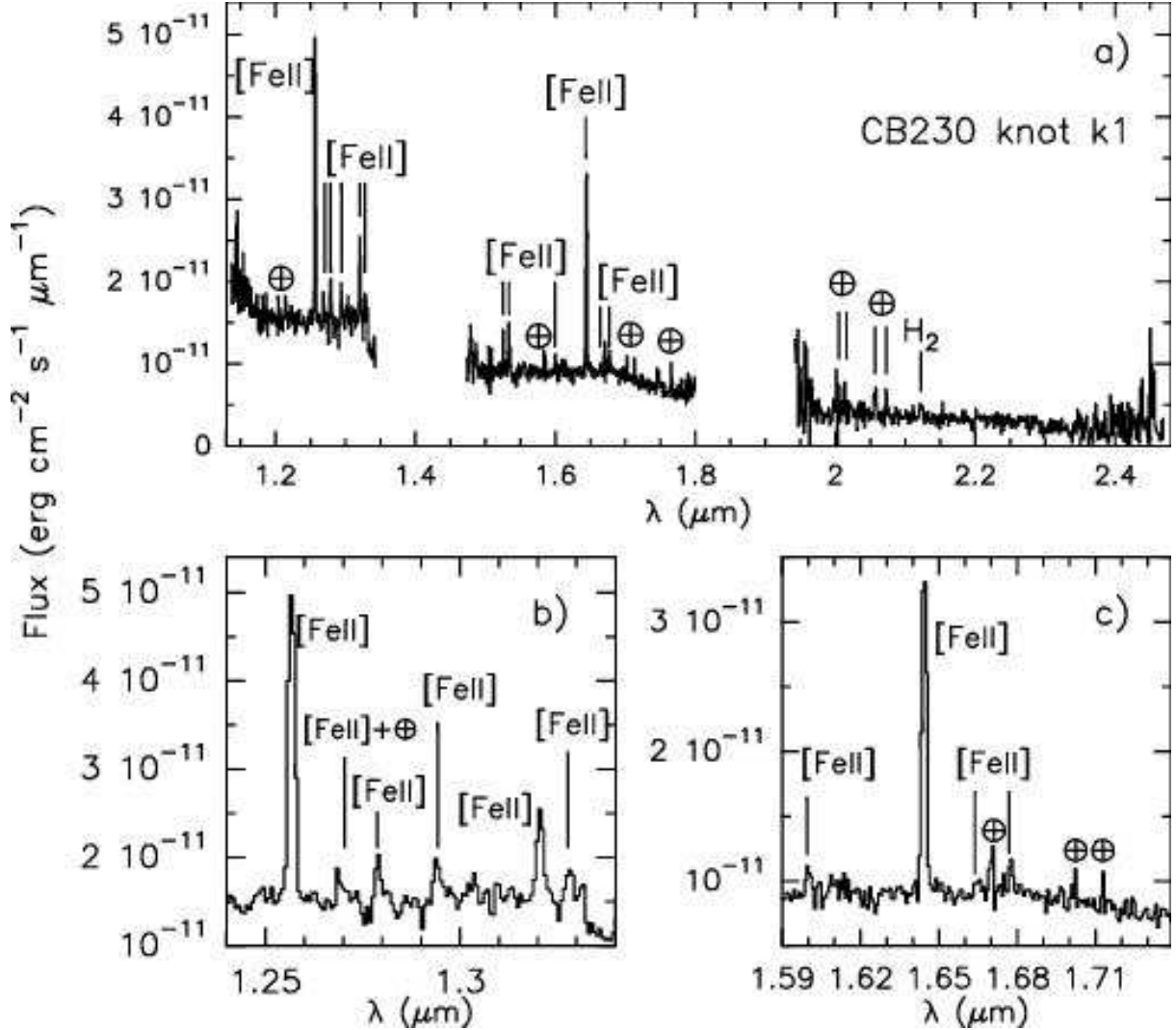}
\caption{
(a) NIR ($JHK$) dereddened spectra ($A_{v}$ = 14~mag)
obtained with the TNG towards the [FeII] 
knot k1 (see Fig.~\ref{Fig:Out}). Telluric lines have been marked, too.
(b) and (c) Zoom-in on the spectral regions around  the bright [FeII] emission 
lines at 1.26 and 1.64~${\mu}$m, where several other weaker [FeII] lines have 
been detected. Table~\ref{emlines} provides the complete list of detections 
and tentative detections of emission lines.}
\label{Fig:Speck1}
\end{figure*}

\begin{figure*}
\centering
\includegraphics[width=16cm]{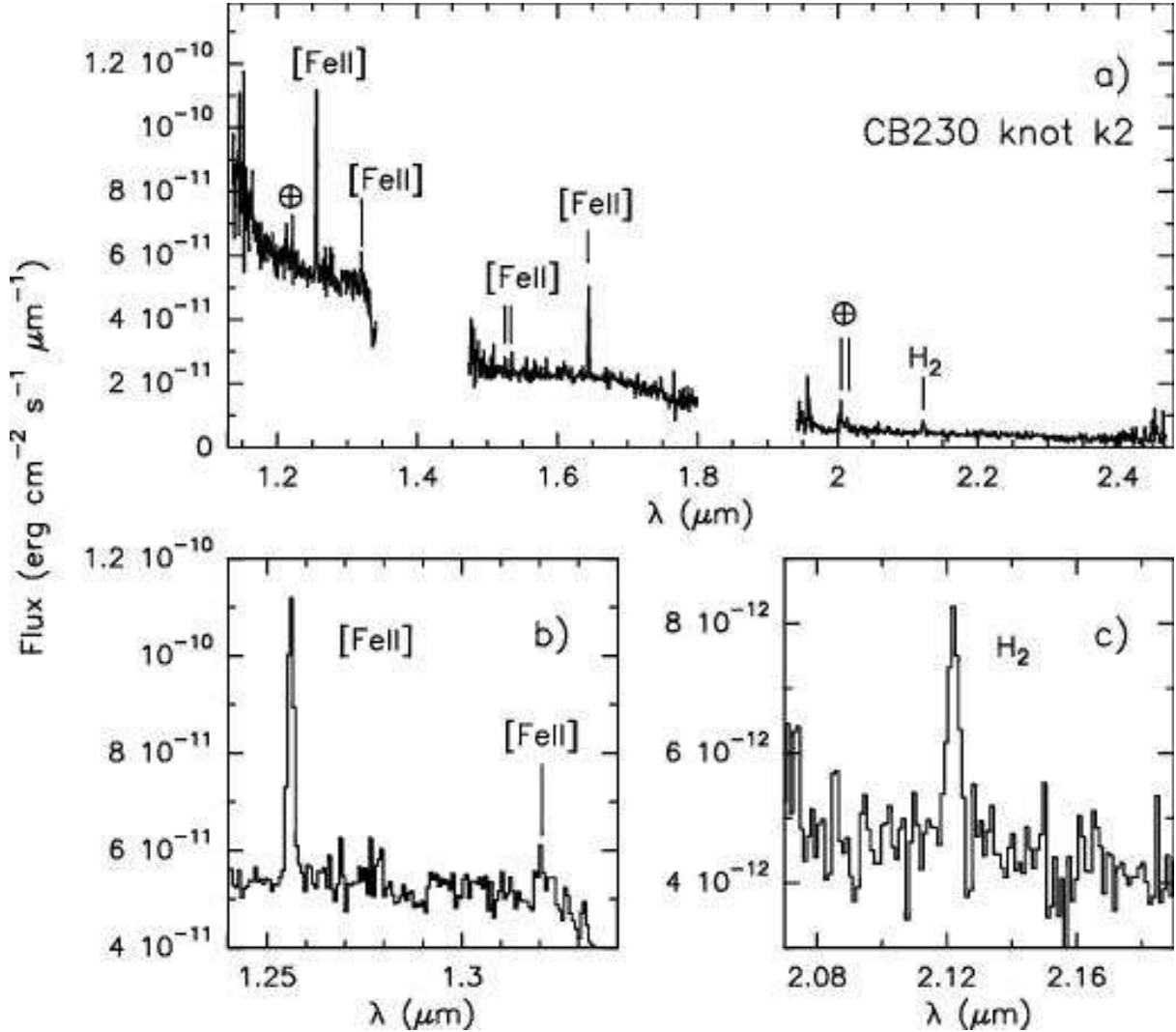}
\caption{
(a) NIR ($JHK$) dereddened spectra ($A_{v}$ = 18.5~mag) 
obtained with the TNG telescope towards 
the [FeII] knot k2 (see Fig.~\ref{Fig:Out}). Telluric lines have been marked, 
too. (b) and (c) Zoom-in on the spectral regions where the bright [FeII] 
emission line at 1.26~${\mu}$m and the H$_{2}$ emission 2.12~${\mu}$m line 
have been detected. Table~\ref{emlines} provides the complete list of 
detections and tentative detections of emission lines.}
\label{Fig:Speck2}
\end{figure*}

\begin{figure*}
\centering
\includegraphics[width=16cm]{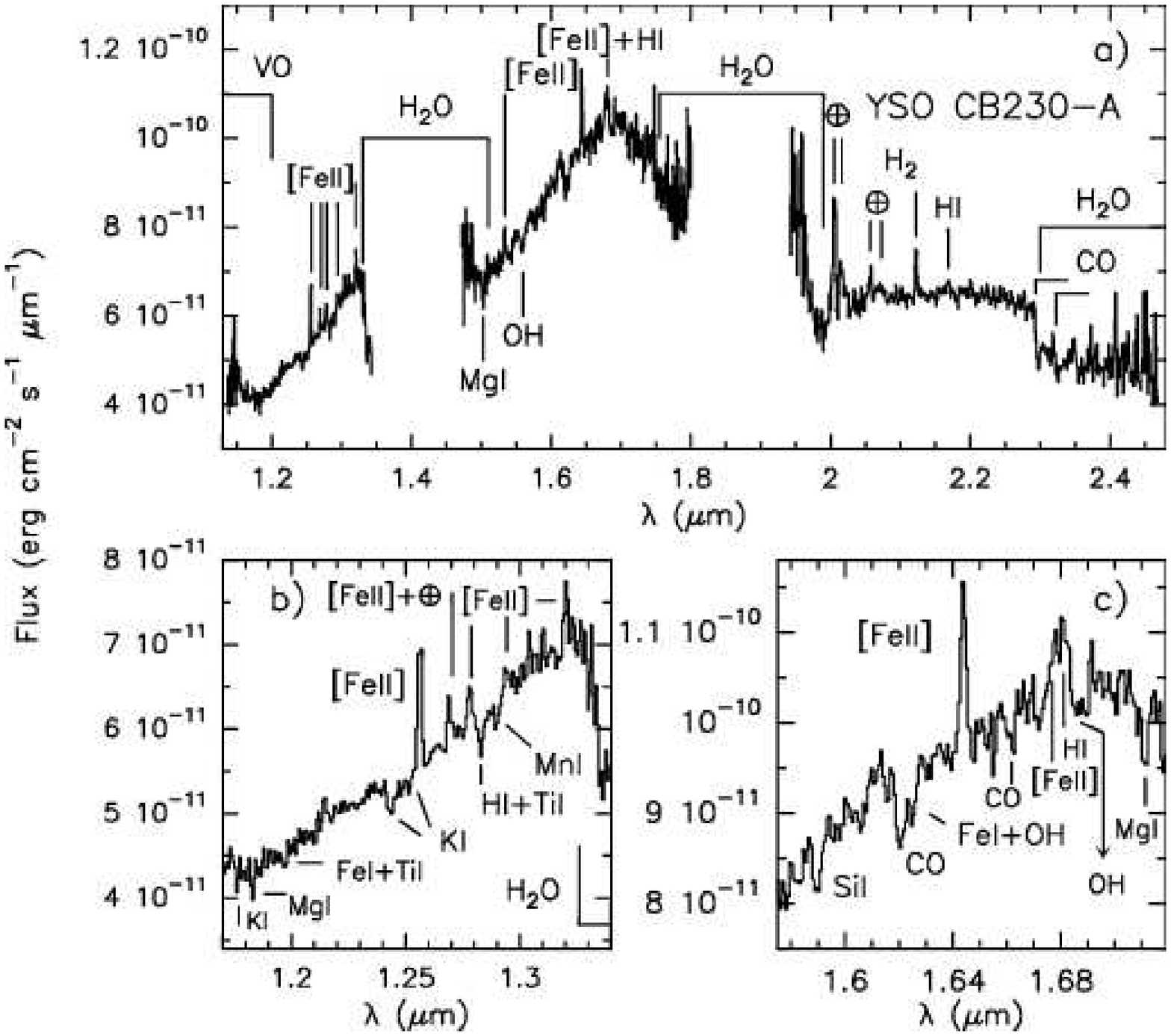}
\caption{
(a) NIR ($JHK$) dereddened spectra ($A_V$ = 8.4~mag) 
acquired by the TNG towards the YSO 
CB230-A, which drives the jet/outflow system (see Fig.~\ref{Fig:Out}). 
Telluric lines are indicated. Tables~\ref{emlines} and \ref{aslines}
provide the complete list of detections and tentative detections of
emission and absorption lines, respectively.
We note the absorption band at wavelengths less than 1.2~$\mu$m, possibly
due to VO,
the three H$_2$O absorption bands around 1.4, 1.9 and 2.4~$\mu$m, as
well as the two CO absorption bands at 2.29 and 2.32~$\mu$m.  
Panels (b) and (c) present a zoom-in view of the spectral regions where the bright [FeII]
emission lines at 1.26 and 1.64~$\mu$m are detected. Note that
for the sake of clarity all the absorption features (detections and tentative 
detections), except for MgI at 1.50~$\mu$m and OH at 1.56~$\mu$m, are 
labelled only in panels (b) and (c).}
\label{Fig:Specyso}
\end{figure*}

The complete $JHK$ spectra of the knots k1 and k2 and of the YSO 
CB230-A are shown in Figs.~\ref{Fig:Speck1}, \ref{Fig:Speck2}, and 
\ref{Fig:Specyso}, respectively. The spectra were dereddened using the 
visual extinctions derived from [FeII] line ratios (see below).
In Table~\ref{emlines}, we list fluxes and identification for all the detected 
emission lines at the three positions (k1, k2, and CB230-A). We note that we 
considered detections to be only those lines with an S/N $\ge$ 3 and with
line widths 
of comparable or larger than the instrumental resolution.
Table~\ref{emlines} also contains a number of tentative detections
(also labelled in Figs.~\ref{Fig:Speck1}, \ref{Fig:Speck2}, and 
\ref{Fig:Specyso}) with a S/N $\simeq$ 2.
The three spectra show a rich set of [FeII] emission lines, as well as
the H$_{2}$ emission line at 2.12~${\mu}$m: in particular, the knot k1 
exhibits a large number of bright emission features. These confirm the 
presence of H$_{2}$
emission and unveil the structure of the [FeII] line-emission,
providing improvements to the interpretations of narrow-band 
filter data by Massi et al.\ (\cite{ma2004}).

\subsection{YSO spectral shape}  

All spectra show an underlying continuum emission, whose shape (before 
correction for extinction) is similar in all directions, i. e. 
towards both the YSO and the knots. However, in the YSO spectrum, the 
continuum emission is more intense and exhibits several emission and 
absorption features. Longward of 1.3~$\mu$m, the beginning of an absorption band 
can be seen in all spectra, which can be identified as being due to water (see 
Sect.~\ref{abso:sect}).
A check of the spectra uncorrected for telluric absorption 
confirms that this band is intrinsic and not a telluric 
feature. 
The continuum emission from the knots most likely originates from the scattering 
of radiation by dust grains. The ubiquitous
presence of this water absorption band indicates that
the illumination source is the YSO. 
Another absorption band clearly visible longward of $2.29$~$\mu$m, is easily
identifiable as the CO(2,0) band. 
This clearly indicates that the central star photosphere and/or the innermost
region of the circumstellar disk are being observed through a cavity produced 
by the outflow.

\section{The YSO}
\subsection{YSO classification}

The classification of both CB230-A and CB230-B+C can be checked
by using Spitzer data to derive their spectral indices between 2 and 24
$\mu$m. Assuming that $\lambda F_{\lambda} \propto \lambda^{\alpha}$, we
estimated that $\alpha \sim 2$ (CB230-A) and $\alpha \sim 2.4$ (CB230-B+C)
from the 2MASS $K$ fluxes and MIPS 24~$\mu$m fluxes. Both
objects 
can be then identified to be Class I sources.

The Mid-Infrared (MIR) colours also, derived from the magnitudes given in
Table~\ref{spitz:tab}, can provide
information about the evolutionary stage of the two objects.
In a [3.6]--[4.5] versus [5.8]--[8.0] colour-colour plot (CCP), both lie
outside the ``disk domain'' (i. e. the plot region dominated by Class II sources) 
from Allen et al.\ (\cite{allen}). CB230-A
colours would be consistent both with those of a reddened photosphere 
(although for an $A_{V}$ of at least 40~mag), as well as with those of an accreting 
protostar (``stage'' I source, according to the classification proposed by
Robitaille et al.\ \cite{robi1}). 
CB230-B+C colours are located in a region occupied mainly by both stage II sources (in 
practise, they are mostly Class II sources) and stage I sources. 
The information from a [3.6]--[5.8] versus [8.0]--[24.0] CCP is
more helpful.
Both CB230-A and CB230-B+C colours are consistent with those of stage I 
sources or heavily reddened ($A_{V} > 20-30$~mag) stage II sources. 

\subsection{YSO SED}

As a further step towards an understanding of the evolutionary stage and 
physical properties of CB230-A, we used a more complete set of 
observational data spanning the wavelength interval from the NIR to the mm. Model 
SEDs could then be fitted to
the data and the relevant physical properties of the YSO derived. 
The source fluxes at different wavelengths, and the effective or estimated
aperture radii for which they were 
measured, are listed in Table~\ref{fluxes:tab}. 

\begin{table}
\caption[]{Fluxes of CB230-A used for SED fitting. Those for CB230-B+C are 
also listed when the two (unresolved) sources are not blended with CB230-A.
\label{fluxes:tab}}
\begin{tabular}{c c c c}
\hline
Band & CB230-A & CB230-B+C & Aperture radius\\
($\mu$m) & (mJy) & (mJy) & ($\arcsec$) \\
\hline
$1.235^{a}$ & $3.702 \pm 0.211$ & $< 0.748$ & 4 \\
$1.662^{a}$ & $21.159$ & $1.128 \pm 0.124$ & 4 \\
$2.159^{a}$ & $42.57$ & $3.893 \pm 0.218$ & 4 \\
$3.6^{b}$ & $64.3 \pm 0.2$ & $2.51 \pm 0.08$ & $2.4$ \\
$4.5^{b}$ & $69.1 \pm 0.4$ & $2.70 \pm 0.09$ &$2.4$ \\
$5.8^{b}$ & $70.4 \pm 0.4$ & $2.9 \pm 0.2$ & $2.4$ \\
$8^{b}$ & $45.4 \pm 0.4$ & $5.27 \pm 0.07$ & $2.4$ \\
$12^{c}$ & \multicolumn{2}{c}{$< 250$} & 67 \\
$24^{d}$ & $521 \pm 2$ & $116 \pm 2$ & 3 \\
$25^{c}$ & \multicolumn{2}{c}{$680 \pm 70$} & 67 \\
$60^{c}$ & \multicolumn{2}{c}{$11750 \pm 700$} & 95 \\
$70^{d}$ &  \multicolumn{2}{c}{$24566 \pm 1500$} & 35 \\
$100^{c}$ & \multicolumn{2}{c}{$33530 \pm 2000$} & 134 \\
$450^{e}$ & \multicolumn{2}{c}{$10430$} & 50 \\
$850^{e}$ & \multicolumn{2}{c}{$2330$} & 20 \\
$1300^{f}$ & \multicolumn{2}{c}{$221 \pm 5$} & 6 \\
\hline
\end{tabular}

$^a$ 2MASS catalogue; $^{b}$ Spitzer/IRAC, this work; $^{c}$ IRAS PSC;
$^{d}$ Spitzer/MIPS, this work; $^{e}$ SCUBA, Brand et al.\ (\cite{brand});
$^f$ IRAM 30-m, Launhardt \& Henning (\cite{launhardt97}); 
\end{table}

By integrating the SED, we found that the ratio of the luminosity
for wavelengths longer than $350$ $\mu$m to the bolometric luminosity is
$\sim 0.01$, which would lead to its classification as a Class 0 source
(Andr\'{e} et al.\ \cite{andre}).
Given also its properties in the NIR to MIR, we can therefore understand 
its classification as a transition Class 0/Class I source. 

\begin{figure}
\centering
\includegraphics[width=8.5cm]{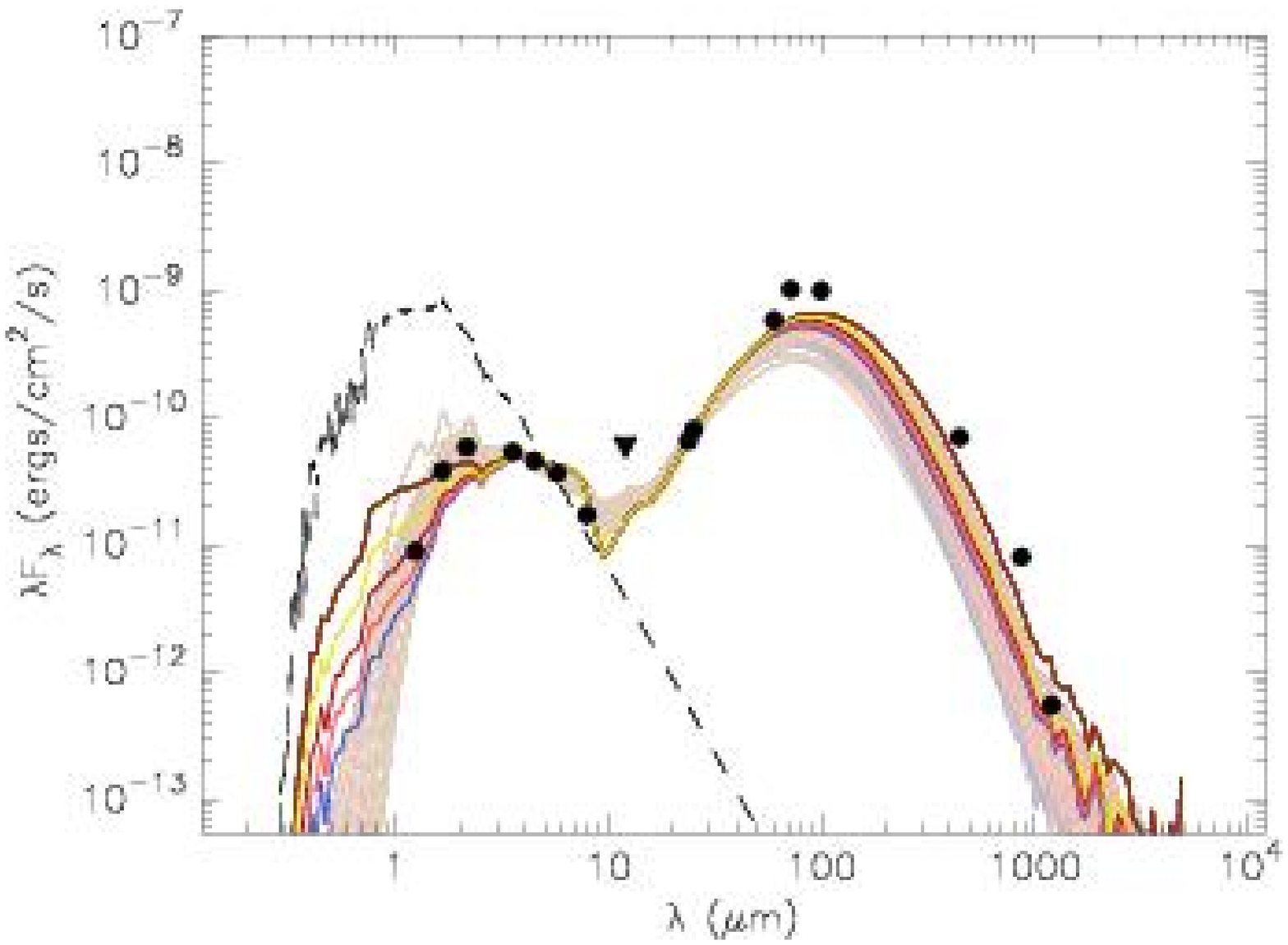}
\caption{Model SEDs (full colour lines in the electronic edition) that best fit the fluxes of CB230-A
(full dots; the full triangle marks an upper limit), within a range
$\chi^{2} - \chi^{2}_{\rm best} < 250$ (per datapoint).}
The dashed line marks the stellar photosphere of the best-fit model.
\label{fig:sed}
\end{figure}

To fit the SED of CB230-A,
we used the online fitting tool by Robitaille et al.\ (\cite{robi2}). 
This exploits a grid of 20000 protostar models for which the radiation 
transfer has been solved, accounting for the central star, envelope and disk and 
including cavities and scattering. A wide range of physical conditions in 
each component is covered (Robitaille et al.\ \cite{robi2}). We constrained 
the distance to be in the range 280--600~pc (see Sect.~\ref{intro}). 
All models (i. e., 16) within a range $\chi^{2} - \chi^{2}_{\rm best} <
250$ (per datapoint) are shown in Fig.~\ref{fig:sed}, where 
$\chi^{2}_{\rm best} = 233.47$ is the value (per datapoint) from the best fit. Clearly, these 
appear to fit the SED reasonably on a by-eye basis. The high values of 
$\chi^{2}$ are due partly to having underestimated the uncertainties
(in most bands, we adopted the photometric errors only). In fact,
we found that by increasing the errors in the IRAC and 1.3 mm fluxes to 
10 \%, 
the $\chi^{2}$ of the best-fit model decreased to 315 (i. e. 20
per datapoint), whereas the range of physical parameters is not significantly
changed. Nevertheless, we also checked that none of the grid models is able
to fit equally well the NIR-MIR and FIR-submm parts of the SED.  
In 
Table~\ref{cb230a:sed}, we list the most relevant physical parameters for the 
best-fit model and their range for the
remaining 15 models within the above $\chi^{2}$ interval.

\begin{table}
\caption[]{Relevant physical parameters of the protostellar model
 that most accurately reproduces the SED of CB230-A and their 
ranges for the remaining 15
 models in the interval $\chi^{2} - \chi^{2}_{\rm best} < 250$ (per datapoint).
\label{cb230a:sed}}
\begin{tabular}{l l l}
\hline
Physical parameter & Best fit & Range \\
\hline
$\chi^{2}$ & 3735.51 & 3804.37--7668.63 \\
Interstellar extinction $A_{V}$ (mag) & $0.00$ & $3.79$--$18.02$ \\
Distance (pc) & 295 & 282--513 \\
Age (yr) & $5.2 \times 10^{4}$ & $2.7 \times 10^{4}$--$10^{5}$ \\
Central mass (M$_{\sun}$) & $0.4$ & $0.15$--$0.5$ \\
Star radius (R$_{\sun}$) & 4 & $3.4$--$4.9$ \\
Star temperature (K) & 3600 & 2908--3759 \\
Accretion rate (M$_{\sun}$ yr$^{-1}$) & $10^{-4}$ & $10^{-4}$--$10^{-6}$ \\
Envelope outer radius (AU) & 5784 & 1968--7757 \\
Cavity aperture angle (deg) & 23 & 21--29 \\
Disk mass (M$_{\sun}$) & $0.03$ & $7.7 \times 10^{-5}$--$0.05$ \\
Disk inner radius (AU) & $0.1$ & $0.06$--$0.3$ \\
Disk accretion rate (10$^{-8}$~M$_{\sun}$ yr$^{-1}$) & $2.3$ &  
                    $0.0077$--$3.6$ \\
Inclination to the line of sight (deg) & $31.8$ & $18.2$--$31.8$ \\
YSO inner extinction $A'_{V}$$^{a}$ (mag) & $48.09$ & $0.09$--$37.33$ \\ 
Total photospheric extinction & & \\
\,\,\,\,\,\,\,\,\,                    $A_{V}+A'_{V}$ (mag) & $48.09$ & $6.5$--$42$ \\
Total system luminosity (L$_{\sun}$) & $2.585$ & $0.846$--$4.287$ \\
Total system luminosity & & \\ 
\,\,\,\,\,\,\,\,\,                    scaled to $d=450$~pc (L$_{\sun}$) & $6$ & $2.2$--$5.7$ \\
\hline
\end{tabular}

$^a$ I. e., the extinction from the outer edge of the circumstellar envelope
to the stellar surface along the line of sight.  
\end{table}

Clearly, a picture emerges of a very young low-mass ($\sim 0.5$~M$_{\sun}$) 
central protostar with a high accretion rate. This agrees with its 
previous classification as a transition Class 0/Class I source. The age 
and mass determination also agree with the compilation by Froebrich (\cite{froeb}).
Its inclination with respect to the line-of-sight 
is not large, as expected since 
only one side of a cavity is visible in the NIR. A circumstellar disk 
is present with a mass $\sim 0.01$~M$_{\sun}$. This is in accord with a disk 
mass $\sim 0.1$~M$_{\sun}$ inferred by Launhardt (\cite{launhardt01}) based 
on interferometric mm observations, since Robitaille et al.\ (\cite{robi2}) 
checked that by using their fitting tool, the agreement between 
fitted disk masses and other disk mass determinations
(when these are available),
is always good to within one order of magnitude. 
However, we note that some of the best-fit models 
exhibit disk masses much smaller than $\sim 0.01$~M$_{\sun}$, 
although the accretion rate is always quite 
high. This implies that even the available MIR data are
unable to constrain strongly the circumstellar disk physical properties. 

Another problem is that all models appear to be underluminous with respect to the 
estimated bolometric luminosity. This partly arises because of the range of 
fitting distances; after scaling all total luminosities instead to a distance of 450~pc (at 
which the reported bolometric luminosity estimates are derived) we achieve closer 
agreement between measurement and predictions of bolometric
luminosity (see Table~\ref{cb230a:sed}). 
Nevertheless, the total luminosity remains underestimated. This is clear 
from Fig.~\ref{fig:sed}, where all model SEDs are always below the datapoints 
in the FIR-to-mm range. This reflects the intrinsic inadequacies of the 
model grid in describing the physical properties of
CB230-A already noted. These models do not account for example for the 
luminosity generated by the envelope accretion (Robitaille et al.\ \cite{robi1}), 
which may in fact be important in this case.

We also attempted to fit the SED of CB230-B+C with the same constraints as 
for CB230-A (but obviously without datapoints in the FIR to mm range). The 
best-fit model has $\chi^{2}_{\rm best} = 17.84$ (per datapoint) and its 
relevant physical parameters are listed in Table~\ref{cb230bc:sed}, along with
those of the models (36) with $\chi^{2} - \chi_{\rm best}^{2} < 30$ (per 
datapoint).
Unfortunately, the lack of measurements in the FIR and submm ranges prevents 
us from constraining strongly many of the physical parameters. The best-fit
model is that of a relatively evolved M$\sim 1.5$~M$_{\sun}$ protostar with an 
edge-on disk that is no longer accreting. 
This can probably be discarded, based on the 
interferometric mm measurements by Launhardt (\cite{launhardt01}) that set an 
upper limit of $0.006$~M$_{\sun}$ to the disk mass of CB230-B+C. However, 
there still remains a combination of Class I and Class II sources with central 
masses in the range $\sim 0.1 - 1.5$~M$_{\sun}$ that are able to fit the SED 
of CB230-B+C. Given the possible binary nature of this source, 
more data are also required to arrive at a more definite identification. This source 
will not be discussed further in the following.

\begin{table}
\caption[]{Relevant physical parameters of the protostellar model
 that most accurately reproduces the SED of CB230-B+C and their ranges for the 
 remaining 36
 models in the interval $\chi^{2} - \chi^{2}_{\rm best} < 30$ (per datapoint).
\label{cb230bc:sed}}
\begin{tabular}{l l l}
\hline
Physical parameter & Best fit & Range \\
\hline
$\chi^{2}$ & 142.72 & 242.36--382.14 \\
Interstellar extinction $A_{V}$ (mag) & $15.04$ & $5.26$--$24.96$ \\
Distance (pc) & 525 & 316--602 \\
Age (yr) & $1.9 \times 10^{6}$ & $5.2 \times 10^{3}$--$10^{7}$ \\
Central mass (M$_{\sun}$) & $1.4$ & $0.10$--$1.59$\\
Star radius (R$_{\sun}$) & 2.4 & $1.6$--$4.3$ \\
Star temperature (K) & 4540 & 2559--5655 \\
Accretion rate (M$_{\sun}$ yr$^{-1}$) & 0 & 0--$4.5 \times 10^{-5}$ \\
Disk mass (M$_{\sun}$) & $0.037$ & $2.7 \times 10^{-5}$--$0.029$ \\
Disk inner radius (AU) & $36$ & $0.044$--$16.6$ \\
Disk accretion rate (10$^{-7}$~M$_{\sun}$ yr$^{-1}$) & $4.0$ &
                    $0.00019$--$1.69$ \\
Inclination to the line of sight (deg) & $87.13$ & $18.2$--$87.13$ \\
YSO inner extinction $A'_{V}$$^{a}$ (mag) & $459.4$ & $4 \times 10^{-5}$--$182$ \\
Total system luminosity (L$_{\sun}$) & $8.230$ & $0.33$--$6.91$ \\
\hline
\end{tabular}

$^a$ I. e., the extinction from the outer edge of the circumstellar envelope
to the stellar surface along the line of sight. 
\end{table}

\subsection{Scattering and extinction in the NIR}
\label{red:scatt}

In principle, the NIR spectrum of the YSO provides a powerful means to 
constrain further 
its total photospheric absorption. The most intense emission lines are
due to three [FeII] transitions at 1.257~$\mu$m, 1.321~$\mu$m, and 1.645~$\mu$m.
Since these [FeII] lines correspond to the same upper energy level
(see Table~\ref{emlines}), the line ratios
1.645/1.321~$\mu$m and 1.645/1.257~$\mu$m can be used to estimate the
extinction towards the [FeII]-emitting regions.
The three lines are within the observed wavelength range of grism $JH$; 
determination of these ratios is therefore straightforward and no intercalibration
that exploits partly overlapping spectral intervals is required. The fluxes were 
computed
from Gaussian fits to the lines and the noise was measured in small
adjoining spectral intervals. Finally, the extinction was derived towards the
YSO and the two knots following Nisini et al.\ (\cite{nisi}), and is listed
in Table~\ref{ext:lr}. The quoted uncertainties were obtained by propagating
the noise. Clearly, there appear to be some discrepancies in the extinction
estimated from the two line ratios. Nisini et al. (\cite{nisi}) pointed out that
these discrepancies are generally found when deriving the extinction from the 
two line ratios, such that the 1.645/1.257-ratio appears to overestimate
$A_{V}$. They recommend using the 1.645/1.321-ratio, ehich they found provides 
results that are more consistent with other independent determinations.
However, in our case it is evident that the two values of $A_{V}$ are always
equal within the uncertainties, that appear then dominated by the
line signal-to-noise ratio, rather than by the uncertainty in the transition
probabilities suggested by Nisini et al. (\cite{nisi}).
However, in Appendix~\ref{ap:b} we propose an alternative explanation for
the disagreement between the extinction values obtained from the two line ratios: 
the 1.257~$\mu$m line intensity can be underestimated 
because of unaccounted-for telluric absorption.
In the following, we follow the prescription of Nisini et al. 
(\cite{nisi}) and use extinctions derived from the 1.645/1.321-ratio.

The error
in line ratios due to inaccurate correction of the atmospheric absorption
(and large-scale differences in system sensitivity) can be estimated to be 
$\sim 10$\%, according to the discussion in Appendix~\ref{ap:a}. 
This would cause the extinction $A_{1.644~{\mu}m}$ to be 
over- or under-estimated 
by $\sim 0.2$~mag, which translates into $\Delta A_{V} \sim 1.2$~mag. Comparing 
this with the values listed in Table~\ref{ext:lr} shows that in all cases but 
one (i. e. k1), the uncertainty in the reddening due to the signal-to-noise 
ratio of the single lines is larger or of the order of that intrinsic to
the telluric correction.

Finally, it is worth noting that from our SCUBA maps
in the sub-mm continuum (Brand et al.\ \cite{brand}), we expect a total
extinction (i. e., over the entire width of the clump)
$A_{\rm V} = 37$~mag, or $A_{\rm V} = 18.5$~mag for objects 
in the centre of the clump. The values inferred from the [FeII] line
ratios towards the knots k1 and k2 are equal or less than that. 
It is also noteworthy that the upper limit to the total
photospheric extinction ($A_{V}+A'_{V}$, where $A_{V}$ is the extinction 
up to the outer edge of 
the circumstellar envelope and $A'_{V}$ that from there 
to the photosphere), listed in Table~\ref{cb230a:sed},
is of the same order as that inferred from the sub-mm emission.

\begin{table}
\caption[]{Visual extinction derived from [FeII] line ratios.
\label{ext:lr}}
\begin{tabular}{c c c}
\hline
Source   & \multicolumn{2}{c}{$A_{V}$} \\
  &  1.645/1.321 & 1.645/1.257 \\
  & (mag)   &  (mag) \\
\hline
k1 & 14.0(1.1) & 14.0(0.2) \\
k2 & 18.5(3.2) & 14.5(3.4) \\ 
CB230-A & 8.4(3.5) & 12.3(1.3) \\
\hline
\end{tabular}
\end{table}

Nevertheless, the extinction derived for the YSO spectrum, although 
within the range of total
photospheric extinction ($A_{V}+A'_{V}$) listed in Table~\ref{cb230a:sed}
and obtained from the SED fits, is lower than $18.5$~mag. This, as well
as the lowest values of $A_{V}+A'_{V}$ obtained, is consistent
with a picture in which the jet has removed part of the gas and dust
from the line of sight, although we would expect a lower extinction from
the knots, as well (see below).
If dereddened by $A_{V} = 8.4$~mag, the intrinsic colour of the YSO is
$J-K = 2.2$, which would be consistent with NIR colours of main sequence dwarf
stars of type later than L2. Their photospheric temperatures are lower than 
the temperatures obtained from the SED fits (see again 
Table~\ref{cb230a:sed}), but their spectral type  
would be consistent with the presence of water and CO absorption in the 
spectrum.
However, to derive the photospheric extinction from the [FeII] lines
we must assume that these lines originate close to the central protostar. 
Instead, there are indications that other effects must also be taken into account.
First of all, many models of those that describe the SED most accurately exhibit a higher
total photospheric extinction. Secondly,  
the continuum appears redder than is usual for
spectra of late-M and early-L giants, 
and spectra of dwarf
stars with strong water and CO absorption bands (see
Sect.~\ref{abso:sect}).
Thirdly, the extinction {\em decreases} from the knots to the YSO: we
would however expect the YSO to be more embedded than the knots, since the knots are
emerging from the cavity. 
And finally, once they are dereddened, 
the $J$ flux density of
knots k1 and k2 is higher than that of the YSO at the shortest wavelengths of band
$J$: this cannot however be true if the continuum radiation from the knots
is produced by dust scattering of infrared emission from the YSO.
In this respect, the extinction derived for the YSO from the 1.645/1.257~$\mu$m ratio
would enable 
a more consistent picture, 
although we remark again
that in the literature this ratio is found to overestimate extinction systematically. 

In particular, the following effects may alter the intrinsic 
[FeII] line ratios:

\begin{enumerate}
\item the extinction derived from the [FeII] line ratios may differ from
the actual reddening of the continuum spectrum of the 
YSO (and the knots) because these
lines 
might trace 
a different environment (the densest part of the jet),
which is only seen in projection towards the YSO (i. e. matter between
this environment and the YSO
could not be accounted for in the extinction derivation);

\item the observed [FeII] line ratios might be affected by 
absorption lines 
in the background photospheric spectrum being almost coincident
with some of the emission lines and, then, lowering 
their measured flux;

\item part or all of the observed radiation from the YSO might have been
scattered by the dust before being extincted.
\end{enumerate}

With reference to item~3, we tried to infer the effects of scattering by 
adopting the results of Whitney et al.\ (\cite{whitney}). Using a 
two-dimensional Monte Carlo radiative transfer code, they modelled NIR 
scattering in cavities, swept out by jets, of the emission from a protostar 
(surrounded 
by a circumstellar disk) embedded in an infalling, rotating cloud with a 
density profile $\sim r^{-2}$. For a number of models, they indicated the ratios 
$R_{\rm {\lambda}}$ of the scattered flux to the intrinsic source flux 
($R_{\rm {\lambda}} = F_{\rm{s,\lambda}} / F_{\rm{0, \lambda}})$ in the 
$J, H, K$, and $L$ bands. We fitted a $\lambda^{-\alpha}$ function to their 
average values for $R_{\lambda}$ ($R_{\rm J}=0.0013$, $R_{\rm H} = 0.0071$, and
$R_{\rm K} = 0.018$) and used this function to correct the spectra for 
scattering approximately. At first we assumed that the observed radiation was first 
scattered and then extincted; we therefore modified the intrinsic line ratios of 
[FeII] by accounting for the derived $\lambda^{-\alpha}$ ``scattering'' 
function. These new line ratios then replaced the intrinsic values in the 
formula that defines $A_{V}$ as a function of the observed [FeII] line ratios. This 
yielded negative values for $A_{V}$, showing either that the used set of 
$R_{\lambda}$ is unrepresentative of the source-cavity system, or that only 
the continuum was scattered, but the [FeII] emission was only
extincted. 

\begin{figure}
\centering
\includegraphics[angle=0,width=8.5cm]{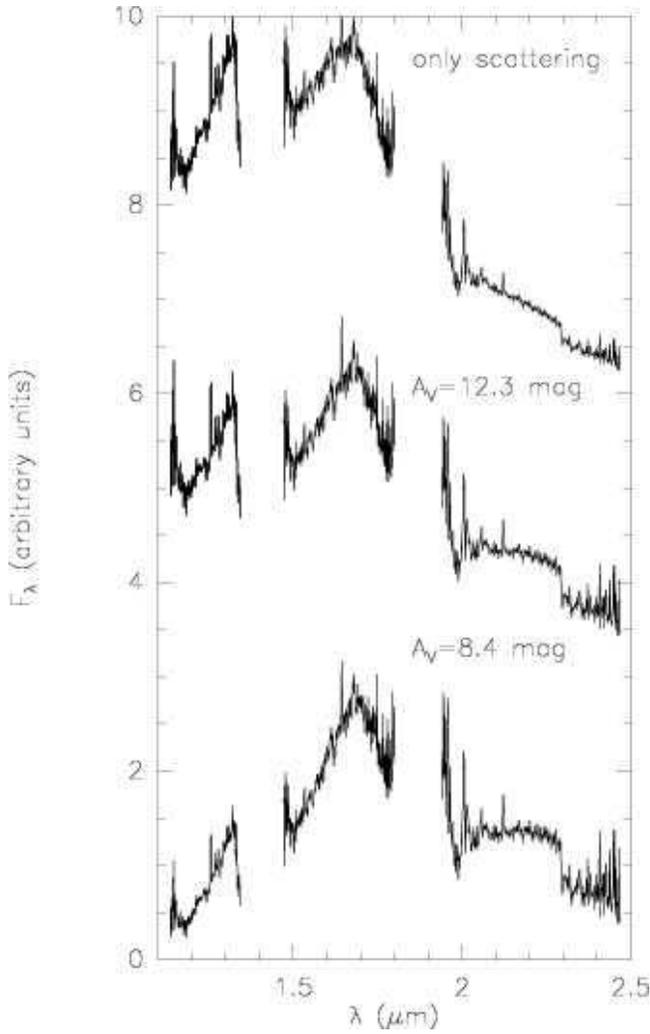}
\caption{From top to bottom: YSO spectrum corrected only for scattering
(see Sect.~\ref{red:scatt}), corrected for $A_{V} = 12.3$~mag and corrected 
for $A_{V} = 8.4$~mag. All spectra have been normalised to the peak of the 
emission and shifted along the $y$-axis. 
\label{Fig:3sp}}
\end{figure}

Next, we assumed that the observed radiation was scattered within the 
cavity but not subsequently extincted. This would be the case if the jet had 
almost completely cleared the cavity and this was inclined towards the 
observer. Using the same models 
that were later used by 
Whitney et 
al.\ (\cite{whitney}), Whitney \& Hartmann (\cite{wh:har}) found
that if the angle between the cavity axis and the line-of-sight is less than 
66$\degr$, the image of the cavity no longer looks bipolar because of the 
extinction of its red side.
In the case of CB230, since the cavity is not bipolar (see Fig.~\ref{Fig:Out}), 
it must be inclined towards the observer, which is also suggested by the 
far smaller red lobe of the CO outflow. It is therefore plausible that scattered 
light from the YSO is 
little extincted 
in its path to the observer.
Remarkably, if we use the $\lambda^{-\alpha}$ function instead of the
reddening law and correct the spectrum only for scattering, we obtain a bluer 
continuum (see Fig.~\ref{Fig:3sp}) and the [FeII] 1.645/1.321 and 1.645/1.257 
line ratios are only 7\% and 14\% less than the intrinsic ones, respectively. 
Then, the adopted scattering correction would naturally
yield the intrinsic [FeII]
line ratios, suggesting that the observed YSO spectrum is also consistent
with pure scattered emission from the innermost regions of the globule, 
emerging through the cavity.


Figure~\ref{Fig:3sp} shows examples of the YSO spectrum after correction for 
scattering only, after correction for dust extinction of $A_{V} = 12.3$~mag, and after 
correction for $A_{V} = 8.4$~mag. 

In summary, our scattering model is oversimplified. Without more 
detailed modelling of CB230-A, it is difficult to disentangle the 
effects of scattering and extinction. We therefore adopt the conservative 
approach of correcting the YSO spectrum only for a reddening of $A_{V} = 8.4$~mag, 
derived from the [FeII] line ratios. This choice does not affect 
the measured absorption-line equivalent widths, or the emission-line fluxes, 
which remain good estimates provided that they are produced 
along the jet, but far from the protostar-disk system. 

\subsection{Absorption features towards CB230-A: a tentative spectral 
identification} \label{abso:sect}

The presence of absorption features in the YSO spectrum, if produced by the photosphere,
would allow us to constrain the physical parameters of the 
central protostar more reliably. 
Spectral identifications of Class I sources have been successfully carried out 
in the NIR by a number of authors (see e. g. Nisini et al.\ \cite{nisia}, 
Doppmann et al.\ \cite{doppmann}, and references therein), but always using 
medium or high spectral resolution data. These authors found ubiquitous CO 
overtone absorption bands, and a number of absorption lines, due
to the ``cold'' photospheres of the central protostars. However, CB230-A is 
the first instance of a Class I source exhibiting both water absorption bands 
and other absorption features at low spectral resolution. 

In Fig.~\ref{Fig:Specyso}, the spectrum towards the YSO is shown after
dereddening by $A_{V} = 8.4$~mag, i. e.\ the value derived from the [FeII]
1.645/1.321 line ratio.
The continuum emission peaks in the $H$ band and there are prominent absorption
bands longward of $1.3$ ${\mu}$m (H$_2$O) and $2.29$ ${\mu}$m (first and second CO
overtone).
Other absorption lines (also labelled in Fig.~\ref{Fig:Specyso}) were 
found in the spectrum; we identified them by comparison with spectra of
late-M and early-L type photospheres. Both detections (S/N $\ge$ 3) and
tentative detections of absorption features are listed in Table~\ref{aslines},
following the spectral catalogues of Wallace et al.\ (\cite{wally}; MK
standards in the $J$-band), Meyer et al.\ (\cite{meyer}; MK standards in the
$H$-band), Lancon \& Rocca-Volmerange (\cite{lancon}) and Dallier et al.\
(\cite{dallier}; both papers discuss NIR spectra of O- to M-stars of
luminosity classes I to V), Reid et al.\ (\cite{reid}) and Cushing et al.\
(\cite{cushing}; both about M- and L-dwarfs). 

\begin{table}
\caption[]{Atomic absorption lines and molecular bands observed towards 
the YSO CB230-A: detections and tentative detections \label{aslines}}
\begin{tabular}{ccc}
\hline
Atom/Molecule & $\lambda$$^a$ & $EW$$^b$ \\ 
              & ($\mu$m)      &  (\AA) \\
\hline
KI         & 1.1778 & 0.67(0.27)$^c$ \\
MgI        & 1.1828 & 0.89(0.27) \\
FeI, TiI   & 1.1974 & 0.64(0.36)$^c$ \\
KI         & 1.2437 & 1.76(0.29) \\
KI         & 1.2529$^d$ & 0.59(0.29)$^c$ \\
HI (Pa$\beta$), TiI    & 1.2827 & 1.25(0.31) \\
MnI        & 1.2905 & 0.70(0.31)$^c$ \\
MgI        & 1.5028 & 2.63(0.62) \\
OH(2--0)P(11) & 1.5575 & 1.08(0.28) \\
SiI$^e$        & 1.5894 & 1.97(0.32) \\
CO(6,3)    & 1.6187  & 3.7(0.4) \\
FeI, OH(2--0)P(15)$^f$ & 1.6250$^f$ & 1.4(0.4) \\
CO(8,5) & 1.6617 & 0.94(0.37)$^c$ \\
OH(2--0)P(18) & 1.6877 & 0.72(0.28)$^c$ \\
MgI & 1.7113 & 1.39(0.32) \\
CO(2,0) & 2.2935 & 12.00(0.83) \\
\hline
\end{tabular}

$^a$ All wavelengths of the atomic transitions are taken from Cushing et 
al.\ (\cite{cushing}), except for that of MgI at 1.18~$\mu$m, taken from the NIST atomic 
spectra database. The OH wavelengths are from Maillard et al.\ (1976), while
those of CO are from Dallier et al.\ (\cite{dallier}) and Meyer et al.\ 
(\cite{meyer}). 
$^b$ The equivalent widths were measured in the
spectra after dereddening by using the visual extinctions 
derived from the [FeII] 1.645/1.321 line ratio (see Tab.~\ref{ext:lr});
however, this should have little effect on the found values.
$^c$ S/N $\sim$ 2: tentative detection.
$^d$ Blended with the [FeII] emission line at 1.2567~$\mu$m 
(see Tab.~\ref{emlines}). 
$^e$ Possible contamination due to the OH(2--0)P(13) band at 1.5902~ $\mu$m.  
$^f$ The wavelengths of FeI and OH(2--0)P(15) are 1.6236 and 1.6265~ $\mu$m,
respectively. Here we report the observed wavelength.
\end{table}

Unfortunately, as can be seen in Table~\ref{aslines}, 
most equivalent widths are only few sigma 
and the $K$-part of the spectrum is of 
too low signal-to-noise to enable reliable detection of any absorption lines. Nevertheless, a few
lines are discernible in the $J$ and $H$ spectra.
Furthermore, a few absorption features were identified beyond any doubt:
the H$_2$O bands at 1.3--1.5, 1.75--2.05, and 2.3--3.2 ${\mu}$m and the
CO(2,0), CO(3,1), and CO(6,3) overtone bands longward of 2.29, 2.32, and 1.62
${\mu}$m, respectively. 


Given the low spectral resolution and low signal-to-noise of the YSO spectrum,
and the uncertainty in the spectral slopes, we decided not to perform a 
spectral
identification by using fits of template spectra. We instead used the identified
features to check their consistency with the photospheric temperatures derived
by the SED fit. These were compared with known features of late-type dwarfs 
(class V) and giant (class III) stars: we must consider the latter stellar
types in particular because the 
radius of the central protostar is expected to be larger, corresponding to a low 
surface gravity, which affects gravity-sensitive lines. Furthermore, the presence 
of a circumstellar disk causes veiling of the absorption lines, which must be 
taken into account. We also applied the diagnostics for late-type star 
identification proposed by Meyer et al.\ (\cite{meyer}) in the $H$-band
and by Wallace et al.\ (\cite{wally}) in the $J$-band, and derived the values of 
a few spectral indices based on the water and CO bands, and Na lines, in 
late-M to L
dwarfs, as devised by Reid et al.\ (\cite{reid}), McLean et al.\ (\cite{mclean}),
and Allers et al.\  (\cite{allers}).

\begin{figure}
\centering
\includegraphics[angle=0,width=8.5cm]{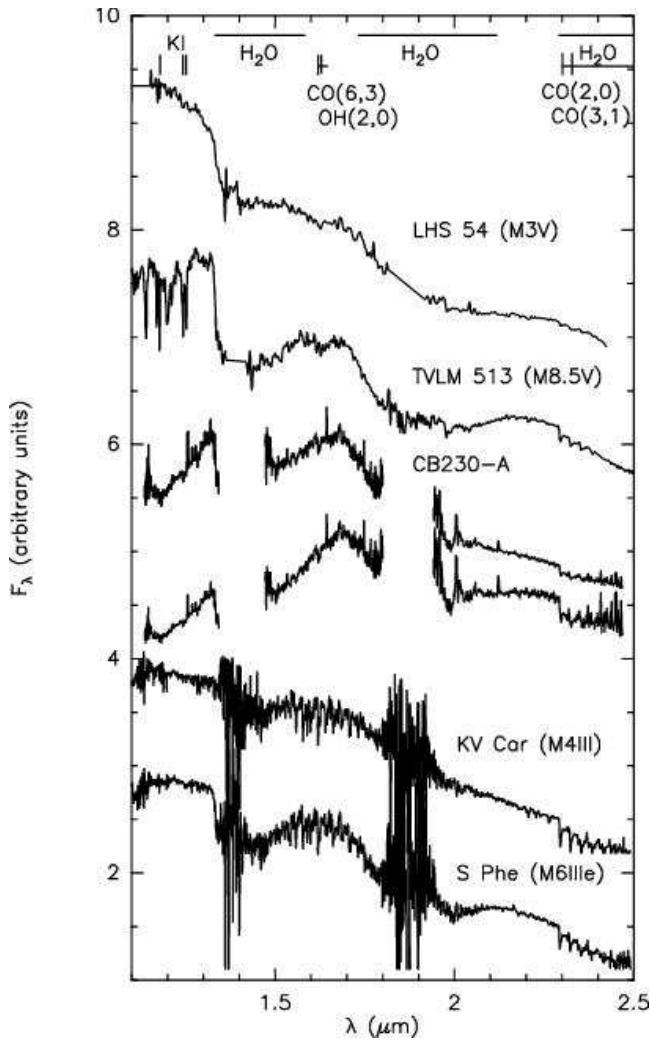}
\caption{NIR spectra of CB230-A (corrected for $A_{V} = 8.4$ mag, upper, and
for scattering only, lower) compared with low-resolution spectra of 
LHS 54 (an M3V star, from Leggett et al.\ 2000), TVLM 513-46546
(an M8.5V star, from Leggett et al.\ 2001), KV Car (an M4III star,
from Lancon et al.\ 2000) and S Phe (an M6IIIe star, from Lancon et al.\ 2000). 
The main lines and spectral features identified in the spectrum of CB230-A are
labelled. Each spectrum has been normalised to its flux at $1.7$ $\mu$m.
\label{Fig:spec-comp}}
\end{figure}

In Fig.~\ref{Fig:spec-comp}, the spectrum of CB230-A (for two
different extinction and scattering corrections) is compared with
low-resolution spectra of M3V, M8.5V, M4III, and M6IIIe stars.
In spite of the uncertainties in the slopes, the H$_{2}$O and 
CO bands ensure that the YSO spectrum resembles that of late M (or even early L dwarfs), 
or late M giants.
Conversely, the absorption lines identified in the $J$ band are more typical
of either early M dwarfs or M giants. The clearest detections are lines from KI,
which are in fact gravity-sensitive (McGovern et al.\ \cite{mcgovern}).
The CO(6,3) band is also a signature of giant (luminosity class III) stars.
However, the measured equivalent widths of all CO band-heads seem 
more typical of K rather than M stars (see also Origlia et al.\
\cite{origlia} for a list of typical values). 
These disagreements probably reflect both
a lower gravity with respect to dwarf stars and veiling by the NIR emission
of the circumstellar disk. Other indications of lower gravity come
from the possible detection of the A--X 0--1 VO band around 1.18~${\mu}$m
and the measured 1.14~${\mu}$m Na-index (as defined by  
Allers et al.\ \cite{allers}). In summary, the YSO spectrum can be
identified as later than M2 (with M2 a conservative limit) with some indications of
a lower gravity than in dwarf stars. 


\subsection{Cold photosphere and/or active accretion disk}

Since the mass of the protostar is expected to be $< 0.5$~M$_{\sun}$ 
(see Table~\ref{cb230a:sed}), which agrees with the mass determinations
($< 0.1$~M$_{\sun}$) of the circumstellar disk,
clearly most of the bolometric luminosity must be generated by
the accretion. 
According to the $T_{\rm eff}$ scale adopted by Meyer et al.\ (\cite{meyer})
and tabulated in their Table~2, 
$T_{\rm eff}$ is $\sim 3530$~K for a luminosity class V
($\sim 3710$~K  for a luminosity class III)
at the upper end of the inferred spectral type interval  (M2).
This is within the range of central temperatures derived from the
SED fits (see Table~\ref{cb230a:sed}).
However, the protostar models of the SED-fitting tool adopt the pms tracks
of Siess et al.\ (\cite{siess}) for the central star, which. 
do not account for accretion. It would therefore be meaningful to compare
our observations with models of accreting protostars.
In this respect, Palla \& Stahler (\cite{ps91}; \cite{ps92}; \cite{ps93}) 
follow the
growth of an accreting protostar up to intermediate masses. Then, the
evolution of the bolometric luminosity (including the fraction provided by
accretion), $T_{\rm eff}$, and protostellar mass can be evaluated along their
birthline.
 
From our CO maps obtained at the JCMT antenna (Brand et al.
\cite{brand}), we estimate an outflowing mass rate of $\sim 6\times
10^{-7}$~M$_{\sun}$\, yr$^{-1}$. Consequently, following theoretical models
of star formation (e.g. Shu et al. \cite{shu}), we expect an accretion rate
10--100 times higher, i.e. around $10^{-5}$~M$_{\sun}$\, yr$^{-1}$.
This is also within the range derived from the SED fits (see 
Table~\ref{cb230a:sed}).
For such a value, Palla \& Stahler (\cite{ps93}) found that a bolometric
luminosity of $\sim 7.7$~L$_{\sun}$ is produced by a protostar of $\sim 
1.5$~M$_{\sun}$ with $T_{\rm eff} \sim 4467$~K, i. e. a more massive
central object hotter than envisaged so far. 
We note that if we use $d$ = 288~pc, i. e. the lower limit of the distance
determination (Sect.~\ref{intro}), we find more precise agreement, although
still marginal, obtaining $T_{\rm eff} $ $\simeq$ 3980~K
and $M \sim 0.6$~M$_{\sun}$.

Another way to improve agreement between the observations and the
model is to increase the accretion rate, which would produce a higher 
bolometric luminosity for a given mass and,
as a consequence, $T_{\rm eff}$ could be lowered. However, if we assumed an
accretion rate of $10^{-4}$~M$_{\sun}$\, yr$^{-1}$ then a protostar of
$< 1$~M$_{\sun}$ would need less than $10^{4}$~yrs to be assembled. But this
would be less than our estimated outflow dynamical age ($\sim 3 \times 
10^{4}$~yrs 
at $d$ = 450~pc, assuming an inclination with respect to the line-of-sight
of $i \sim 45^{\degr}$). If the dynamical age can be considered to be a lower limit 
to the age of the driving source, this rules out an accretion rate 
of more than $10^{-4}$~M$_{\sun}$\, yr$^{-1}$. 
This again agrees with the upper limit of the accretion rate derived by the
SED fits.

Wuchterl \& Tscharnuter (\cite{wutscha}) presented evolutionary tracks of protostars
following the collapse of Bonnor-Ebert spheres. Their tracks showed that $T_{\rm eff}
\sim 3000$ K and bolometric luminosities of $\sim 7.7$~L$_{\sun}$ can be obtained
for central protostars of $\sim 0.5$~M$_{\sun}$ and $< 0.3$ Myr old. By using
these tracks, we therefore obtain a far closer agreement with our observations. Nevertheless,
according to this model,
the protostar would have already accreted $\sim 90$ \% of the initial clump mass, which is
inconsistent with the high current envelope mass that we have measured
($\sim 3.4$~M$_{\sun}$). 

One other possibility to be considered is that a few spectral features
(namely the strong 
absorption bands of CO and H$_2$O) are {\it not} photospheric in 
origin, but are produced in an active accretion disk. The NIR spectra of 
FU~Ori stars exhibit strong absorption bands of water and first and second 
overtones of CO, in contrast with their optical identification as F--G 
supergiants (Mould et al.\ \cite{mould}). This was 
attributed to the NIR part of FU Ori spectra being generated by an
accretion disk (Calvet et al.\ \cite{calvet91}). Calvet et al.\
(\cite{calvet91b}) showed that for central stars with $T_{\rm eff} < 6000$~K,
mass accretion rates of $>10^{-7}$~M$_{\sun}$\, yr$^{-1}$ are able to
turn the CO bands into absorption. 
Even if the mass of the protostar were 
1~M$_{\sun}$, it could have been accumulated in less than $10^{7}$~yr with an accretion
rate of only $>10^{-7}$~M$_{\sun}$\, yr$^{-1}$. These values lie at the upper 
end of the acceptable ranges of both age and mass. 
It is therefore possible that
at least some of the features in the YSO NIR spectrum are produced in an
active accretion disk rather than in a protostellar photosphere. 

Clearly, these issues will be resolved by future high-resolution 
high-sensitivity NIR spectral observations, which will enable a more 
robust spectral classification to be achieved and allow us to recognise the 
kinematical signature of a disk in the line profiles, irrespective of 
reddening and scattering corrections. 

\subsection{Spectral indicators of the accretion rate}

Muzerolle et al.\ (\cite{muze}) found a tight correlation between the emission
luminosity of Br$\gamma$ and the accretion luminosity in a sample of classical
T-Tauri stars. We can use their Eq.~2 and the line luminosity estimated from 
the flux listed in Tab.~\ref{emlines} (assuming $d$ = 450 pc) to derive an 
accretion luminosity of $\sim 0.15$~L$_{\sun}$. This would be consistent with
``low'' accretion rates ($\sim 10^{-7}$~M$_{\sun}$ yr$^{-1}$) and central 
masses ($\sim 0.1$~M$_{\sun}$), but in total disagreement with the 
observed ``high'' bolometric luminosity, which in this case should be 
produced mostly by accretion. By raising the central mass to 
obtain higher stellar luminosities, we must however resort to the active disk 
scenario to produce the CO bands in absorption, which would again disagree with the 
low accretion luminosity derived from the Br$\gamma$ flux. Part of the 
discrepancy may be due to the likely underestimate of the extinction (or the 
neglect of scattering) already discussed. For instance, an underestimate in $A_{V}$ 
of 3~mag would increase the derived accretion luminosity to $\sim 
4.5$~L$_{\sun}$, in far closer agreement with the estimated bolometric 
luminosity. However, this would assume that the Muzerolle et al. (\cite{muze}) relation 
is applicable in addition to
less evolved YSOs.

\section{The outflow}

\subsection{Physical properties of the jet from the NIR spectra}

The extinction of the knots was derived in Sect.~\ref{red:scatt}
using the ratios of the intense emission lines 
due to three [FeII] transitions at 1.257~$\mu$m, 1.321~$\mu$m, and 1.645~$\mu$m.
The values obtained are listed in Table~\ref{ext:lr} and
we adopted those derived from the 1.645/1.321~$\mu$m ratio.
All the other observed [FeII] lines (see Table~\ref{emlines} and 
Figs.~\ref{Fig:Speck1} and
\ref{Fig:Speck2}) are associated with similar excitation
levels ($\sim$ 11000-12000~K) and therefore cannot be used to derive
temperature estimates (e.g. Nisini et al. \cite{nisi}).
On the other hand, these [FeII] lines have different critical densities
ranging between 10$^4$ and 10$^5$~cm$^{-3}$. Following the model 
developed by Nisini et al. (\cite{nisi}), it is possible to derive an estimate of
the electron density ($n_{\rm e}$) 
by using intensity ratios involving
the brightest transitions which are density-sensitive (i. e. 1.645/1.600, 
1.645/1.533, and 1.645/1.677). For the CB230-A position, we obtain
1.645/1.600 $\ge$ 2.2, 1.645/1.533 = 2.2, and 1.645/1.677 = 4.3, which are 
consistent with $n_{\rm e}$ $\ge$ 10$^5$~cm$^{-3}$. In other words,
the density towards the YSO is higher than the critical density of the [FeII] 
lines, which are therefore expected to be thermalised. For the knot k1, we derived 
1.645/1.600 = 10.3, 1.645/1.533 = 4.3, and 1.645/1.677 = 12.5, which correspond to 
electron densities in the $6 \times 10^3 - 1 \times 10^4$~cm$^{-3}$ range. 
Finally, the 
weaker k2 spectrum indicates 1.645/1.600 $\ge$ 7.5, 1.645/1.533 = 5.4, and 
1.645/1.677 $\ge$ 7.5, again suggesting densities of around 10$^4$~cm$^{-3}$.
Our measurements confirm the findings of Nisini et al. (\cite{nisi})
in HH-flows associated with three SFRs, i. e. that the NIR [FeII] emission
lines trace either high density post-shock portions of the ionised gas or 
regions of a high degree of ionisation and then a high electron density.
Unfortunately, to derive the hydrogen ionisation fraction $x_{\rm
e}$, following the procedure introduced by Bacciotti \& Eisl\"offel
(\cite{bacciotti99}), we would need measurements of oxygen and nitrogen lines 
in the optical and to assume that these lines trace the same 
material that emits the NIR [FeII] lines. To derive a rough estimate of the total 
hydrogen density, $n_{\rm H}$ = $n_{\rm e}$/$x_{\rm e}$, we can only assume that the
ionisation fraction is equal to the typical value measured in a number of jets (e.g. Podio et al. 
\cite{podio}, and references therein), $x_{\rm e}$ = 0.03--0.6.
If this is also valid in the CB230 case, we infer densities $n_{\rm H}$ 
towards the k1-k2 jet of between 10$^4$ and $3 \times 10^5$ cm$^{-3}$.

Finally, the comparison between the [FeII] 1.64~$\mu$m and the H$_2$ 
2.12~$\mu$m intensities indicates that these emission lines are not 
spatially
correlated, in agreement with the scenario in which [FeII] and H$_2$ are excited 
by different mechanisms, and the [FeII] emission originates in dissociative 
shocks. Furthermore, the H$_{2}$ line intensity clearly increases from the outermost
knot to the YSO (see Table~\ref{emlines}).

%
\subsection{The outflow driven by the jet}

Regarding the mass-loss process, it would be instructive to compare
the momentum flux of the atomic jet ($\dot{P}_{\rm jet}$) with that of the 
larger-scale CO outflow ($\dot{P}_{\rm CO}$ $\simeq$ 
10$^{-5}$~M$_{\sun}$\, yr$^{-1}$~km\, s$^{-1}$).
Unfortunately, the present NIR data do not allow us to measure the 
momentum flux of the jet directly. 
By using our derived values of ionisation density
(i. e. in the range $6 \times 10^3 - 1 \times 10^4$~cm$^{-3}$) 
and jet size
($FWHM = 1\farcs2$), and assuming typical values for 
the ionisation fraction of 0.03--0.6, and a jet velocity 
of 100-500~km\, s$^{-1}$, we 
obtained a range of values for $\dot{P}_{\rm jet}$ of
10$^{-5}$--10$^{-2}$~M$_{\sun}$~yr$^{-1}$~km\, s$^{-1}$.
The upper limit to this range is clearly unrealistic, which supports
a high ionisation fraction, $x_{\rm e}$ $\ge$ 0.1, to 
limit the jet momentum flux arbitrarily to the more reasonable range of 
10$^{-5}$--10$^{-4}$~M$_{\sun}$~yr$^{-1}$~km\, s$^{-1}$ (e.g. Podio et al. 
\cite{podio}). In any case, it appears that  
$\dot{P}_{\rm jet}$ $\ge$ $\dot{P}_{\rm CO}$, which indicates that the
[FeII] jet is indeed able to drive the molecular outflow.

\section{Summary and conclusions}

We have presented our study of
the star-forming site located in the globule CB230 
using spectroscopic NIR observations.
We obtained $JH$ and $HK$ spectra of the knots in the [FeII] jet, which 
originates in the low-mass YSO CB230-A, and the protostar itself.
This allowed us to derive the physical properties of the region where 
the jet is launched from. 
We have also retrieved Spitzer archive data in the range 3.6--70~$\mu$m
that, complemented with our NIR spectra, enabled us to constrain the relevant
physical parameters of the protostar.

Our main results are the following:

\begin{enumerate}

\item the spectra of the jet knots show a large number of emission lines, 
including 
a rich set of [FeII] lines. The brightest [FeII] and H$_2$ emission lines are 
spatially uncorrelated, confirming that [FeII] and H$_2$ are excited by different 
mechanisms, in agreement with the models in which [FeII] traces dissociative 
J-shocks and molecular hydrogen traces slower C-shocks.

\item the YSO spectrum exhibits a large number of atomic and molecular 
emission, and absorption features. The characteristics of this spectrum, if 
photospheric in origin, would correspond to a spectral type later than M2 for 
CB230-A. They also suggest a lower gravity than for class V stars, which is 
expected in a 
protostar where the radius is larger than that of a corresponding main 
sequence star. 

\item complementing flux measurements at different wavelengths with the Spitzer
data, we obtained a SED that can be fitted by a young low-mass ($M < 
0.5$~M$_{\sun}$) protostar with a high 
($10^{-4} - 10^{-6}$~M$_{\sun}$\, yr$^{-1}$) accretion rate
and a circumstellar disk with $M \le 0.05$~M$_{\sun}$. These limits agree with 
the constraints obtained from the NIR spectrum and the other available mm and 
sub-mm
observations. The occurrence of a cavity of relatively
small inclination with respect to the line-of-sight is in accordance with the
NIR morphology of the object and explains why a spectrum of the innermost
protostellar region has been obtained.

\item as for CB230-B(,C), the lack of resolved FIR and sub-mm observations
prevented strong constraints by SED fitting being achievable; a large spread in
the central mass ($0.1 - 1.5$~M$_{\sun}$) and evolutionary stages
(from Class I to Class II) was obtained.

\item the absorption bands of water and CO in the YSO spectrum might
  be the signature of an active accretion disk, rather than an indication of
a photospheric origin. NIR spectra of both high spectral resolution
and high signal-to-noise ratio are needed to disentangle the 
photospheric emission from that of the circumstellar disk, further
constraining the protostellar physical parameters.

\item by using intensity ratios involving density-sensitive [FeII] 
emission lines and following the model by Nisini et al. (\cite{nisi}),
we estimated the electron densities of the [FeII] knots along the jet,
obtaining values in the range $6 \times 10^3 - 1 \times 10^4$~cm$^{-3}$.

\item the [FeII] jet appears to be able to drive the larger-scale 
outflow, exhibiting a higher momentum flux than the CO outflow. 

\end{enumerate}

\begin{acknowledgements}
This work is
based on observations made with the Italian Telescopio Nazionale Galileo
(TNG) operated on the island of La Palma by the Centro Galileo Galilei of
the INAF (Istituto Nazionale di Astrofisica) at the Spanish Observatorio del
Roque de los Muchachos of the Instituto de Astrofisica de Canarias.

This work is based in part on observations made with the Spitzer Space Telescope, 
which is operated by the Jet Propulsion Laboratory, California Institute of Technology 
under a contract with NASA.

This publication makes use of data products from the Two Micron All Sky 
Survey, which is a  joint project of the University of Massachusetts and the 
Infrared Processing and Analysis Center/California Institute of Technology, 
funded by the National Aeronautics and Space Administration and the National 
Science Foundation.

This research has also made use of the SIMBAD database, operated at CDS, 
Strasbourg, France.

We wish to thank F. Palla, L. Testi, F. Bacciotti, L. Podio, A. Isella and A. Natta for helpful discussions 
and suggestions.
We also thank an anonymous referee whose comments helped us to improve 
the quality of this work.
\end{acknowledgements}

\begin{appendix}
\section{Estimate of errors due to inaccuracies in the correction
of the atmospheric absorption}
\label{ap:a}

The effect of an inaccurate correction of atmospheric absorption
is assessed as follows. We assume that $T(\lambda,t)$ is
the atmospheric transmission at wavelength $\lambda$ and time $t$.
If the source spectrum $S(\lambda)$ taken at time $t_{1}$ is corrected
with a $T$ estimated at time $t_{2}$, then the ``corrected'' spectrum
will be:

\begin{equation}
S_{\rm est}(\lambda) = S(\lambda) \times \frac{T(\lambda,t_{1})}{T(\lambda,t_{2})}
\end{equation}

The estimated ratio between two lines at wavelengths $\lambda_{2}$ and
$\lambda_{1}$ will then be:

\begin{equation}
\label{fund:1}
R_{\rm est} = \frac{S(\lambda_{2})}{S(\lambda_{1})}
\frac{T(\lambda_{2},t_{1})}{T(\lambda_{2},t_{2})}
\frac{T(\lambda_{1},t_{2})}{T(\lambda_{1},t_{1})}
\end{equation}

\noindent
Irrespective of the shape of $T$, we can then write

\begin{equation}
\frac{T(\lambda_{2},t_{1})}{T(\lambda_{2},t_{2})} =
\frac{T(\lambda_{1},t_{1})}{T(\lambda_{1},t_{2})} + \Delta
\end{equation}

\noindent
so that Eq.~\ref{fund:1} becomes:

\begin{equation}
\label{fund:2}
R_{\rm est} = \frac{S(\lambda_{2})}{S(\lambda_{1})}
(1 + \Delta / \frac{T(\lambda_{1},t_{1})}{T(\lambda_{1},t_{2})})
\end{equation}

\noindent
or

\begin{equation}
\label{fund:3}
\frac{\Delta R}{R} = \Delta/\frac{T(\lambda_{1},t_{1})}{T(\lambda_{1},t_{2})}
= A
\end{equation}

The error in the line ratios should then depend only on the ratio of $T$ at a given
time to that at a different time.

This ratio can be easily estimated from the ratio of (extracted)
spectra of a single object (be it the target or a telluric standard)
taken at different times and, obviously, not yet corrected for the
atmospheric absorption. We note that this ratio also includes 
variations due to large-scale sensitivity differences in the detector area and 
slit misalignments during AB cycles.

As for our observations, we consider spectra of telluric standard stars
(HIP97033 and HIP108772), taken both 7 hours before and just prior to the 
observations of the target. 
The different ratios of $T$ depend almost linearly 
on $\lambda$ in the
relevant intervals within the $J-, H-,$ and $K$-bands, increasing or decreasing
with wavelength. The larger slope (in absolute value) is always found in the $J$-band; it is 
much smaller for the other bands.
We estimated the ratios of $T$ at 1.644, 1.32 and 1.25~$\mu$m
($JH$), and at 2.12 and 1.644~$\mu$m ($HK$).
The quantity $A$ derived by using spectra of telluric standards
within a same ABBA-cycle (time differences of a few minutes) is almost
always $\sim 0.1$, or even less. It is likely that, in this case, $A$
is dominated by variations due to large-scale sensitivity differences and slit
misalignments during the AB-cycles. When using the target spectra
(time differences of a few to 10~minutes), we found again that $A \sim 0.1$. Finally,
using telluric standard spectra separated by $\sim 7$~hrs, we obtained
$A \sim 0.8-1$ for $JH$ and $A < 0.35$ for $HK$. By roughly assuming that
$A$ increases linearly with time, we again found that $A \sim 0.1$ in $JH$ 
(much less in $HK$) within $\sim 1$~hour, i. e. the
maximum time interval between telluric and target observations in $JH$.
Regarding atmospheric absorption, we conclude that line ratios
for data acquired using a same grism should be accurate to within 10\%,
after taking account of both variations due to large-scale
sensitivity differences and slit misalignments during AB-cycles.

Problems arise only in observed wavelength ranges affected by
the strong telluric absorption between the $J$- , $H$- and 
$K$-bands, which correspond to the edges of the three bands. In this respect,
the changes in slopes occurring at wavelengths close to the edges of the three 
segments of spectrum (where the noise also increases) have to be regarded 
with caution. They might have been increased by variations of 
telluric absorption that have been unaccounted for.

\section{Reliability of the [FeII] 1.645/1.257 $\mu$m line ratio
as an extinction estimator}
\label{ap:b}

As stated in the text, the [FeII] line ratios 
1.645/1.321~$\mu$m and 1.645/1.257~$\mu$m can be used to estimate the
extinction towards the [FeII]-emitting regions, since all of these lines originate in 
the same upper energy level. However, Nisini et al.\ (\cite{nisi}) found
that the 1.645/1.257-ratio systematically underestimates the extinction
towards a number of sources. They suggested that this may be due to a
poor determination of the Einstein coefficients. 
This can be checked easily because any discrepancies in the adopted Einstein coefficients
should produce a fixed systematic difference between the values of $A_{V}$ 
determined from the two line ratios for any observational dataset. 
If this difference was then found not to be constant (within errors), 
alternative sources of contamination should be considered.

Such a check is beyond the scope of this paper. We propose
another possibility 
that does not involve the Einstein coefficients; the cause would be
the same as that envisioned by Bailey et al.\ (\cite{basi}), i. e. an error
in the atmospheric correction occurring in low (and possibly also mid) 
spectral-resolution spectra, in this case affecting the 1.257~$\mu$m line.  
This line is in a wavelength range containing the $^{1}\Delta_{g}$
absorption band of O$_{2}$. 
At high spectral resolution, Fig.~8 of Bailey et al.\ (\cite{basi})
shows that there are many narrow lines that lower the 
transmission to almost zero at their central wavelengths, in the same region 
in which the 1.257~$\mu$m line is located. At low resolution,
these high-opacity absorption lines are smoothed away 
and produce a broad
depression in the empirically-obtained atmospheric transmission curve.
If the observed wavelenght range of the [FeII] line corresponded
to one of the narrow, atmospheric  absorption lines,
it would be underestimated following telluric correction.
This would lead to an overestimate of the 1.645/1.257-ratio and, in turn, to 
an overestimate of the extinction when this was derived from the ratio,
the situation which has indeed been found. 
It is difficult to assess the magnitude of the error without a detailed modelling of both
the atmospheric absorption and the [FeII] line. 

\end{appendix}


\begin{thebibliography}{}

\bibitem[2004]{allen} Allen, L.\ E., Calvet, N., D'Alessio, P., et al. \ 2004,
ApJS, 154, 363
\bibitem[2007]{allers}
Allers, K.\ N., Jaffe, D.\ T., Luhman, K.\ L., et al.\ 2007, \apj, 657, 511
\bibitem[2000]{andre}
Andr\'{e}, P., Ward-Thompson, D., \& Barsony, M.\ 2000, From Prestellar Cores to Protostars: 
the Initial Conditions of Star Formation,
in Protostars and Planets IV, ed. V. Mannings, A. P. Boss, \& S. S. Russell 
(Tucson: Univ. Arizona Press), 59
\bibitem[1999]{bacciotti99}
Bacciotti, F., \& Eisl\"offel, J.\ 1999, \aap, 342, 717
\bibitem[2007]{basi}
Bailey, J., Simpson, A., \& Crisp, D.\ 2007, \pasp, 119, 228
\bibitem[1947]{bok}
Bok, B.\ J., \& Reilly, E.\ F.\ 1947, \apj, 105, 255
\bibitem[2008]{brand}
Brand, J., Wouterloot, J.\ G.\ A., Codella, C., \& Massi, F.\ 2008, in preparation 
\bibitem[1991a]{calvet91}
Calvet, N., Hartmann, L., \& Kenyon, S.\ J.\ 1991a, \apj, 383, 752
\bibitem[1991b]{calvet91b}
Calvet, N., Patino, A., Magris, G.\ C., \& D'Alessio, P.\ 
1991b, \apj, 380, 617
\bibitem[2007]{chen} Chen, X., Launhardt, R., \& Henning, T.\ 2007, \apj, 669, 1058
\bibitem[1988]{clemens}
Clemens, D.\ P., \& Barvainis, R.\ 1988, \apjs, 68, 257
\bibitem[2006]{codella}
Codella, C., Brand, J., Massi, F., Wouterloot, J.\ G.\ A., \& Davis, G.\ R.\ 2006, 
\aap, 457, 891
\bibitem[2005]{cushing}
Cushing, M.\ C., Rayner, J.\ T., \& Vacca, W.\ D.\ 2005, \apj, 623, 1115
\bibitem[1996]{dallier}
Dallier, R., Boisson, C., \& Joly, M.\ 1996, \apjs, 116, 239
\bibitem[2005]{doppmann}
Doppmann, G.\ W., Greene, T.\ P., Covey, K.\ R., \& Lada, C.\ J.\ 2005, \aj, 130, 1145
\bibitem[2005]{froeb} Froebrich, D.\ 2005, \apjs, 156, 169
\bibitem[2004]{fazio1} Fazio, G.\ G., Hora, J.\ L., Allen, L.\ E., et al.\
2004, \apj, 154, 10
\bibitem[2007]{gordon} Gordon, K.\ D., Engelbracht, C.\ W., Fadda, D., et al.\
2007, \pasp, 119, 1019
\bibitem[1999]{huard99}
Huard, T.\ L., Sandell, G., \& Weintraub, D.\ A.\ 1999, \apj, 526, 833
\bibitem[1998]{kun}
Kun M.\ 1998, \apjs\ 115, 59
\bibitem[1992]{lancon} Lancon, A., \& Rocca-Volmerange, B.\ 1992, \aaps, 96, 593
\bibitem[2000]{lan000} Lancon, A., \& Wood, P.\ R.\ 2000, \aaps, 146, 217
\bibitem[1997]{launhardt97}
Launhardt, R., \& Henning, Th.\ 1997, \aap, 326, 329
\bibitem[2001]{launhardt01}
Launhardt, R.\ 2001, Fragmentation of a Protostellar Core: The Case of CB 230, in
The Formation of Binary Stars, ed.\ 
H. Zinnecker, \& R.\ D.\ Mathieu, Proceedings of IAU Symp.\ 200, 117 
\bibitem[2000]{leg2000} Leggett, S.\ K., Allard, F., Dahn, C., et al.\
2000, \apj, 535, 965
\bibitem[2001]{leg2001} Leggett, S.\ K., Allard, F., Geballe, T.\ R., Hauschildt, P.\ H.,
\& Schweitzer, A.\ 2001, \apj, 548, 908 
\bibitem[1976]{maillard}
Maillard, J.\ P., Chauville, J., \& Mantz, A.\ W.\ 1976, J.\ of Molec.\ Spectr., 63, 120
\bibitem[1996]{maio} 
Maiolino, R., Rieke, G.\ H., \& Rieke, M.\ J.\ 1996, \aj, 111, 537
\bibitem[2004]{ma2004}
Massi, F., Codella, C., \& Brand, J.\ 2004, \aap, 419, 241
\bibitem[2004]{mcgovern}
McGovern, M.\ R., Kirkpatrick, J.\ D., McLean, I.\ S., et al.\ 2004,
\apj, 600, 1020
\bibitem[2003]{mclean}
McLean, I.\ S., McGovern, M.\ R., Burgasser, A.\ J., et al.\ 2003, \apj, 596, 561
\bibitem[1998]{meyer} Meyer, M.\ R., Edwards, S., Hinkle, K.\ H., \& Strom S.\
E.\ 1998, \apj, 508, 397
\bibitem[1978]{mould}
Mould, J.\ R., Hall, D.\ N.\ B., Ridgway, S.\ T., Hintzen, P., \&
Aaronson, M.\ 1978, \apj, 222, L123
\bibitem[1998]{muze} Muzerolle, J., Hartmann, L., \& Calvet, N.\ 1998, \aj, 116, 2965
\bibitem[2005a]{nisia} 
Nisini, B., Antoniucci, S., Giannini, T., \& Lorenzetti, D.\ 2005a, \aap, 429, 543
\bibitem[2005b]{nisi} 
Nisini, B., Bacciotti, F., Giannini, T., et al.\ 2005b, \aap, 441, 159
\bibitem[1993]{origlia}
Origlia, L., Moorwood, A.\ F.\ M., \& Oliva, E.\ 1993, \aap, 280, 536
\bibitem[1991]{ps91} Palla, F., \& Stahler, S.\ W.\ 1991, \apj, 375, 288
\bibitem[1992]{ps92} Palla, F., \& Stahler, S.\ W.\ 1992, \apj, 392, 667
\bibitem[1993]{ps93} Palla, F., \& Stahler, S.\ W.\ 1993, \apj, 418, 414
\bibitem[2006]{podio}
Podio, L., Bacciotti, F., Nisini, B., et al.\ 2006, \aap, 456, 189
\bibitem[2001]{reid} Reid, I.\ N., Burgasser, A.\ J., Cruz, K.\ L.,
Kirkpatrick, J.\ D., \& Gizis, J.\ E.\ 2001, \aj, 121, 1710
\bibitem[2005]{reach} Reach, W.\ T., Megeath, S.\ T., Cohen, M., et al.\
2005, \pasp, 117, 978
\bibitem[1985]{r&l}
Rieke, G.\ H., \& Lebofsky, M.\ J.\ 1985, \apj, 288, 618
\bibitem[2004]{rieke04} Rieke, G.\ H., Young, E.\ T., Engelbracht, C.\ W.,
et al.\ 2004, \apjs, 154, 25
\bibitem[2006]{robi1} Robitaille, T.\ P., Whitney, B.\ A., Indebetouw, R., Wood, K., \&
         Denzmore, P.\ 2006, \apjs, 167, 256
\bibitem[2007]{robi2} Robitaille, T.\ P., Whitney, B.\ A., Indebetouw, R.,  \& Wood, K.\ 
2007, \apjs, 169, 328
\bibitem[2000]{shu}
Shu, F.\ H., Najita, J.\ R., Shang, H., \& Li, Z.-Y.\ 2000, X-Winds Theory and Observations,
in Protostar and Planets IV,
ed. V. Mannings, A.\ P. Boss, \& S.\ S.\ Russel (Tucson: University of Arizona Press), 789
\bibitem[2000]{siess} Siess, L., Dufour, E., \& Forestini, M.\ 2000, \aap, 358, 593
\bibitem[1992]{straizys}
Straizys, V., Cernis, K., Kazlauskas, A., \& Meistas, E.\
1992, Baltic Astr., 1, 149 
\bibitem[2000]{wally} 
Wallace, L., Meyer, M.\ R., Hinkle, K., \& Edwards, S.\ 2000, \apj, 535, 325
\bibitem[1993]{wh:har}
Whitney, B.\ A., \& Hartmann, L.\ 1993, \apj, 402, 605
\bibitem[1997]{whitney} 
Whitney, B.\ A., Kenyon, S.\ J., \& G\'{o}mez, M.\ 1997, \apj, 485, 703
\bibitem[2003]{wutscha}
Wuchterl, G., \& Tscharnuter, W.\ M.\ 2003, \aap, 398, 1081
\bibitem[2003]{young}
Young, C.\ H., Shirley, Y.\ L., Evans, N.\ J.\ II, \& Rawlings, J.\ M.\ C.\ 2003, 
\apjs, 145, 111
\bibitem[2006]{yy06}
Young, C.\ H., Bourke, T.\ L., Young, K.\ E., et al.\ 2006,
\aj, 132, 1998
\bibitem[1992]{yuncle92}
Yun, J.\ L., \& Clemens, D.\ P.\ 1992, \apj, 385, L21
\bibitem[1994a]{yuncle94a}
Yun, J.\ L., \& Clemens, D.\ P.\ 1994a, \apjs, 92, 145
\bibitem[1994b]{yuncle94b}
Yun, J.\ L., \& Clemens, D.\ P.\ 1994b, \aj, 108, 612
\end{thebibliography}
\end{document}